\documentclass[]{aa}  

\usepackage{graphicx}
\usepackage{txfonts}
\usepackage{amsmath}
\usepackage{adjustbox}
\usepackage{hyperref}
\hypersetup{colorlinks=true,allcolors=[rgb]{0,0,0.5}}
\usepackage[dvipsnames]{xcolor}
\usepackage[version=4]{mhchem} 
\usepackage{soul}

\makeatletter
\renewcommand*\aa@pageof{, page \thepage{} of \pageref*{LastPage}}
\makeatother

\newcommand{\refree}[1]{#1}

\begin{document} 

         \title{Under the magnifying glass: A combined 3D model applied to cloudy warm Saturn type exoplanets around M-dwarfs}
         
   \titlerunning{Under the magnifying glass}

   \author{S.~Kiefer \inst{1, 2, 3}
            \and
            N.~Bach-M{\o}ller \inst{2, 3, 4}
            \and
            D.~Samra \inst{2}
            \and
            D.~A.~Lewis \inst{2,3}
            \and
            A.~D.~Schneider \inst{1, 4}
            \and 
            F.~Amadio \inst{4,1}
            \and
            H.~Lecoq-Molinos \inst{2, 3, 1}
            \and
            L.~Carone \inst{2}
            \and
            L.~Decin \inst{1}
            \and
            U.~G.~J{\o}rgensen \inst{4}
            \and
            Ch.~Helling \inst{2,3}
            }

   \institute{Institute of Astronomy, KU Leuven, Celestijnenlaan 200D, 3001 Leuven, Belgium\\
              \email{sven.kiefer@kuleuven.be}
         \and
              Space Research Institute, Austrian Academy of Sciences, Schmiedlstrasse 6, A-8042 Graz, Austria
         \and
             Institute for Theoretical Physics and Computational Physics, Graz University of Technology, Petersgasse 16 8010 Graz
         \and
              Centre for ExoLife Sciences, Niels Bohr Institute, {\O}ster Voldgade 5, 1350 Copenhagen, Denmark
             }

   \date{Received ...; accepted ...}

  \abstract
   {Warm Saturn type exoplanets orbiting M-dwarfs are particularly suitable for in-depth cloud characterisation through transmission spectroscopy due to their favourable stellar to planetary radius contrast. The global temperatures of warm Saturns suggest efficient cloud formation in their atmospheres which in return affects the temperature, velocity, and chemical structure. However, modelling the formation processes of cloud particles consistently within the 3D atmosphere remains computationally challenging. 
   }
   {The aim is to explore the combined atmospheric and micro-physical cloud structure, and the kinetic gas-phase chemistry of warm Saturn-like exoplanets in the irradiation field of an M-dwarf. The combined modelling approach will support the interpretation of observational data from present (e.g. JWST, CHEOPS) and future missions (PLATO, Ariel, HWO).
   }
   {A combined 3D cloudy atmosphere model for HATS-6b is constructed by iteratively executing the 3D General Circulation Model (GCM) \texttt{expeRT/MITgcm} and a detailed, kinetic cloud formation model, each in its full complexity. Resulting cloud particle number densities, particles sizes, and material compositions are used to derive the local cloud opacity which is then utilised in the next GCM iteration. The disequilibrium H/C/O/N gas-phase chemistry is calculated for each iteration to assess the resulting transmission spectrum in post-processing.
   }
   {The first model atmosphere that iteratively combines cloud formation and 3D GCM simulation is presented and applied to the warm Saturn HATS-6b. The cloud opacity feedback causes a temperature inversion at the sub-stellar point and at the evening terminator at gas pressures higher than 10$^{-2}$~bar. Furthermore, clouds cool the atmosphere between $10^{-2}$ bar and 10 bar, and narrow the equatorial wind jet. The transmission spectrum shows muted gas-phase absorption and a cloud particle silicate feature at $\sim 10\mu$m.
   }
   {The combined atmosphere-cloud model retains the full physical complexity of each component and therefore enables a detailed physical interpretation with JWST NIRSpec and MIRI LRS observational accuracy. The model shows that warm Saturn type exoplanets around M-dwarfs are ideal candidates to search for limb asymmetries in clouds and chemistry, identify cloud particle composition by observing their spectral features, and identify the cloud-induced strong thermal inversion that arises on these planets specifically.
   }
   
   \keywords{Planets and satellites: individual: HATS-6b -- Planets and satellites: atmospheres -- Methods: numerical -- Planets and satellites: gaseous planets -- Techniques: spectroscopic
             }

   \maketitle
%

\section{Introduction}
\label{sec:Introduction}

\begin{figure}
        \centering
        \includegraphics[width=\hsize]{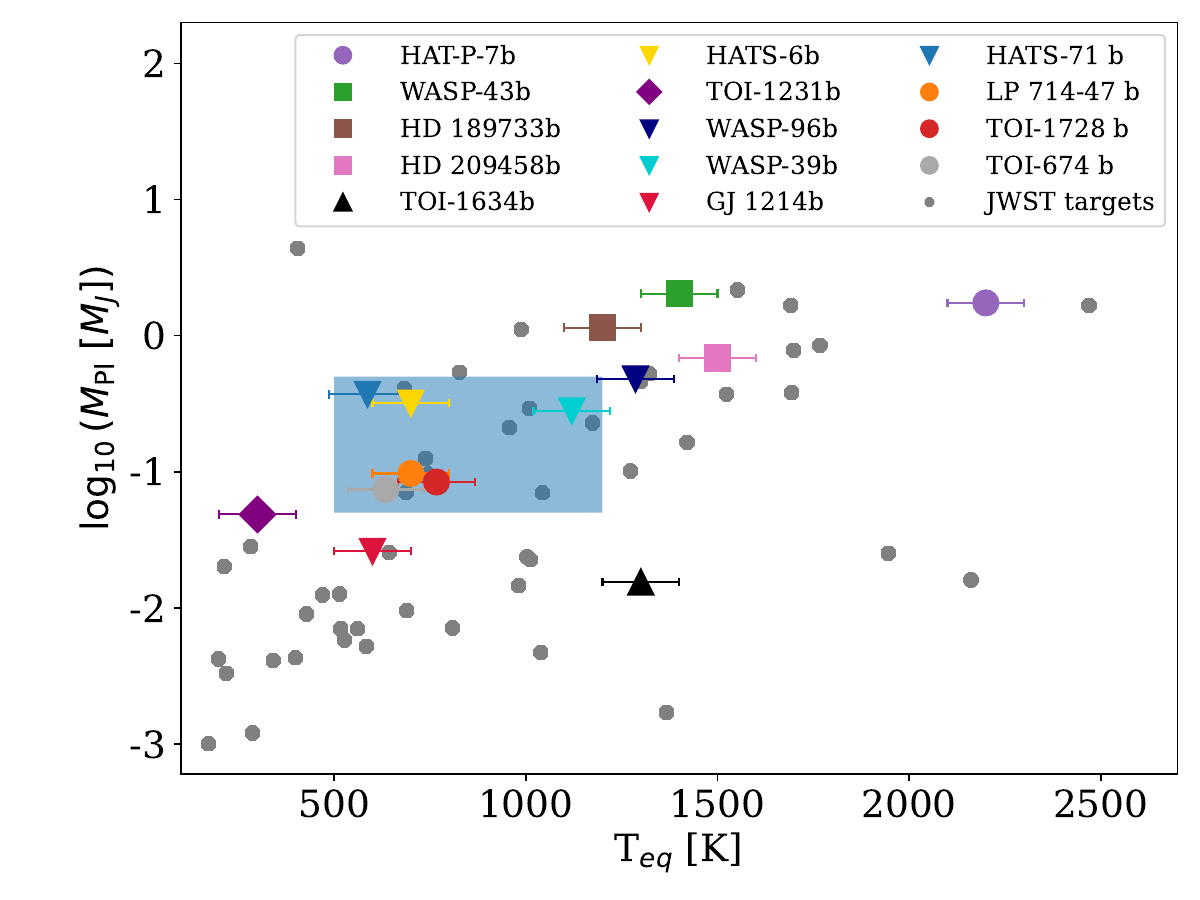}
        \caption{Known gas-giants and JWST targets. The blue shaded area represents the approximate parameter space of warm Saturns. Several often studied exoplanets are shown.}
        \label{fig:overview}
    \end{figure}

    Warm Saturns are a class of Saturn-sized gas-giants with equilibrium temperatures $T_\mathrm{eq}$ between 500 K and 1200 K (Fig. \ref{fig:overview}). The atmospheres of warm Saturns have so far been difficult to characterise because observations with the Hubble Space Telescope (HST) showed either absent or muted spectral features. Such observations can be interpreted either with a cloud-free high metallicity composition or a global extended cloud coverage with lower metallicity \citep{komacek_clouds_2020, carone_indications_2021, wong_hubble_2022}. James Webb Space Telescope (JWST) observations together with better models can resolve the ambiguous results for warm Saturns (or warm Neptunes like WASP~107b). 
    
    The majority of warm Saturns have been observed around metal rich F, G, or K stars \citep{buchhave_jupiter_2018}. A selected few have now been detected around M-dwarfs \citep{canas_toi-3714_2022, lin_unusual_2023,hartman_hats-6b_2015}. Atmosphere characterisation of warm Saturns in the era of JWST provides an excellent opportunity for in-detail cloud and chemistry characterisation across very different host stars. M-dwarfs are too faint for planetary atmosphere characterisation in the optical and around 1~$\mu$m with HST. Their luminosities peak at longer wavelengths, making warm Saturns around M-dwarfs together with their favourable stellar to planetary radius ratio ideal targets for infrared transmission spectroscopy with JWST. 
    
   First studies with cloudless General Circulation Models (GCMs) predict more uniform temperatures of warm Saturns \citep{christie_impact_2022, helling_exoplanet_2023} compared to (ultra) hot Jupiters. Moreover, due to their relatively low planetary global temperature, a very efficient horizontal heat circulation has been inferred \citep[e.g.][]{kataria_atmospheric_2016, komacek_atmospheric_2016}. Thus, a global and highly mixed composition cloud coverage can be expected for warm Saturn type exoplanets \citep{christie_impact_2022, helling_exoplanet_2023}. Clouds scatter light at the top of the planet \citep{rowe_very_2008} leading to less absorbed incoming stellar light, which tends to cool a planetary atmosphere. At the same time, clouds absorb and re-emit outgoing thermal radiation of the planet and thus impose an additional greenhouse effect that can warm the underlying atmosphere. Understanding which effects determine the thermodynamic environment of the planet requires 3D modelling of the cloud properties and their horizontal and vertical distribution, as has been demonstrated for hot Jupiters \citep{lines_simulating_2018, powell_transit_2019, parmentier_cloudy_2021, lee_mini-chemical_2023} and rocky exoplanets \citep{yang_strong_2014, turbet_day-night_2021}. 
   Recent JWST observations, as well as recent theoretical work, showcase that the properties of cloud particles within a planetary atmosphere are not necessarily uniform, even on planets where global cloud cover is expected. In the case of the hot Jupiter WASP-39b, cloud asymmetries between the morning and evening terminator have been predicted \citep{carone_wasp-39b_2023, arfaux_coupling_2024} and confirmed \citep{espinoza_inhomogeneous_2024}.

    For the in-detail characterisation of cloud properties, gas-phase chemistry, and feedback with the 3D wind flow in exoplanet atmospheres, complex 3D cloud models are needed. Clouds and their formation in gas-giant exoplanets have been described with multiple levels of theory. Simplified descriptions assume phase equilibrium to determine where clouds can potentially be present \citep[e.g.][]{demory_inference_2013, webber_effect_2015, crossfield_observations_2015, kempton_observational_2017, roman_modeling_2017, roman_clouds_2021}.  To be able to make predictions of the material composition, cloud particle sizes, location of formation, and effect on the gas-phase abundances by depletion, fully self-consistent, micro-physical theories are needed \citep[e.g.][]{woitke_dust_2003, helling_dust_2006, helling_modelling_2013, powell_formation_2018, woitke_dust_2020, gao_aerosol_2020}. Atmosphere models combining radiative transfer and micro-physical cloud formation are often solved for one dimension \citep{helling_consistent_2008, witte_dust_2011, juncher_self-consistent_2017, gao_aerosol_2020}. However, atmospheric processes such as equatorial wind jets \citep[e.g.][]{showman_atmospheric_2002}, day-night cold traps \citep[e.g.][]{parmentier_3d_2013, pelletier_vanadium_2023}, day-night asymmetry \citep[e.g.][]{perez-becker_atmospheric_2013, komacek_atmospheric_2017, helling_sparkling_2019} and patchy clouds \citep[e.g.][]{line_influence_2016, tan_global_atmospheric_2021} cannot be captured by 1D models.

    3D GCMs are successfully used to understand the cloud-free wind flow and basic thermodynamic structures of extrasolar planets. These models predict the hot spot offset within hot Jupiters due to equatorial superrotation \citep{showman_atmospheric_2002, showman_equatorial_2011, carone_equatorial_2020} which has been observed since \citet{knutson_map_2007}. To model the interaction of clouds and climate, cloud models and 3D GCMs need to be combined, which is computationally intensive. There are currently several approaches to tackle this problem. In a hierarchical approach, the output of GCMs can be used to post-process cloud structures \citep{helling_mineral_2016, kataria_atmospheric_2016, parmentier_transitions_2016, helling_understanding_2019, helling_cloud_2021, robbins-blanch_cloudy_2022, savel_no_2022, helling_exoplanet_2023}. While these models can account for detailed micro-physical cloud formation, they are missing the feedback of clouds on the GCM. So far, only a few models couple micro-physical cloud models with GCMs \citep{lee_dynamic_2016, lee_dynamic_2017, lines_simulating_2018, lines_exonephology_2018}. To reduce the computational cost, some models use simplifications for the atmosphere models \citep{ormel_arcis_2019, min_arcis_2020} or the cloud modelling \citep{christie_impact_2021,lee_mini-chemical_2023}. Others use radiatively passive \citep{parmentier_3d_2013, charnay_3d_2015-1, komacek_vertical_2019} or active \citep{charnay_3d_2015} cloud particle tracers. Others parameterise the advection of clouds and use condensation curves to prescribe the location of clouds in the 3D GCM \citep{parmentier_transitions_2016, parmentier_thermal_2018, tan_effects_2017, tan_global_atmospheric_2021, roman_modeling_2017, roman_modeled_2019, harada_signatures_2021, christie_impact_2021, christie_impact_2022}. In all cases, there is a trade off between computational cost and the detail by which the models describe the interaction of clouds and climate.
    
    Warm Saturns around M-dwarf stars are interesting targets to study cloud formation and chemistry. The existence of such planets question formation models that predict that only more massive host stars produce a protoplanetary disk with enough material to form a gas-giant \citep[e.g.][]{pascucci_steeper_2016}. M-dwarfs have a higher stellar activity than solar-type stars \citep{mignon_characterisation_2023}. Specifically, the higher magnetic activity of M-dwarfs is expected to expose planets around these stars to higher amounts of stellar energetic particles \citep[SEPs, e.g.][]{fraschetti_stellar_2019, rodgers-lee_energetic_2023}, which can affect the chemistry on these planets \citep{venot_influence_2016, barth_moves_2021, konings_impact_2022}.

    HATS-6b is a rare case of a transiting warm Saturn that orbits an M-dwarf host star. HATS-6b was discovered by the HATSouth survey in 2015 and has well-constrained planetary and stellar parameters. The planet has a mass of $0.319 \pm 0.070$ M$_J$, a radius of $0.998 \pm 0.019$ R$_J$ and an orbital period of $3.3253$ d, where M$_J$ is the mass of Jupiter and R$_J$ the radius of Jupiter. The zero-albedo equilibrium temperature is $\approx 700$ K \citep{hartman_hats-6b_2015}. The host star, HATS-6, is an early M-dwarf (M1V) with a mass of $0.57$ M$_\odot$, a radius of $0.57$ R$_\odot$, a metallicity of [Fe/H] = $0.2 \pm 0.09$, and an effective temperature of $T_\mathrm{eff} = 3724$~K, where M$_\odot$ is the mass of the Sun and R$_\odot$ is the radius of the Sun. Consequently, HATS-6b has one of the deepest transit depths known with $(R_P/R_\star)^2 = 0.0323 \pm 0.0003$ around 600~nm, where $R_P$ is the observed planetary radius and $R_\star$ the host star's radius. In addition, the host star has a J band magnitude of 12.05, making HATS-6b very well suitable for atmospheric characterisation in the infrared. HATS-6b, together with eight other warm Saturns, have been selected as targets for two general observer programs for the James Webb Space Telescope (JWST) Cycle 2 (GO~3171 and 3731).

    This paper has two aims. Firstly, we aim to explore the atmospheric, micro-physical cloud and gas-phase structure of the warm Saturn HATS-6b orbiting an M-dwarf. We use a combined model in the form of step-wise iterations between a detailed cloud formation description and \texttt{expeRT/MITgcm}, a 3D GCM with full radiative transfer and deep atmosphere extension. The second goal is to demonstrate how the combined modelling approach can help to support the interpretation of the data from space missions (e.g. JWST) for warm Saturn type planets. The step-wise iterative approach between the GCM (Sect.~\ref{sec:Model_Mitgcm}) and cloud structure (Sect.~\ref{sec:Model_clouds_static}) is described in Sect.~\ref{sec:Model}. The evaluation of the combined model for the warm Saturn HATS-6b is presented in Sect.~\ref{sec:res}. The resulting atmospheric solution and transmission spectra of HATS-6b are shown in Sect.~\ref{sec:sim}. The discussion is in Sect.~\ref{sec:Discussion} and the conclusion in Sect.~\ref{sec:conclusion}.

\section{Approach}
\label{sec:Model}

    The step-wise iterative process applied in this study is initiated by 3D GCM simulations of the temperature, gas, densities, and wind velocities (Sect.~\ref{sec:Model_Mitgcm}). 1D profiles are extracted from this 3D atmosphere solution and used as input for a kinetic cloud formation model and the local cloud opacities are calculated (Sect.~\ref{sec:Model_clouds_static}). The cloud opacity structure is then used as input for the 3D GCM and a new atmosphere structure is calculated. Each of the models is executed in their full complexity. Such an iteration is possible because the cloud formation time scales are considerably shorter than the hydrodynamic timescales \citep{helling_dust_2001, helling_dust_2006, powell_formation_2018, kiefer_fully_2024}. The stopping criterion for the iteration is described in Sect.~\ref{sec:Model_stopping}. In order to explore how the iterative processes may affect the spectral information, transmission spectra are created for each iteration. This is done by first modelling the non-equilibrium gas-phase chemistry of H/C/N/O species based on the final output from each GCM run (Sect.~\ref{sec:Model_argo}). The resulting relative concentrations of the gas species are then used to calculate the transmission spectra for each iteration and the JWST's observability of the differences is assessed (Sect.~ \ref{sec:Model_clouds_trans})

    \subsection{3D atmosphere}
    \label{sec:Model_Mitgcm}    
    
    The 3D temperature and horizontal (zonal and meridional) wind velocities in the atmosphere of HATS-6b are simulated with the non-gray 3D GCM \texttt{expeRT/MITgcm} \citep{schneider_exploring_2022, carone_equatorial_2020}. This model solves the hydrostatic primitive equations (HPE) in vertical pressure coordinates on a rotating sphere  \citep[see, e.g.][]{showman_atmospheric_2009}. Hydrostatic equilibrium and the ideal gas law as equation of state are assumed. The model uses the dynamical core of \texttt{MITgcm} \citep{adcroft_implementation_2004} which solves the HPE on an Arakawa C type cubed-sphere grid. Here, we employ the nominal horizontal resolution C32\footnote{C32 corresponds to a resolution of 128x64 in longitude and latitude} and a vertical grid with 41 logarithmically spaced grid cells between $1\times 10^{-5}$~bar and 100 bar and six linearly spaced grid cells between 100 bar and 700 bar. \texttt{expeRT/MITgcm} couples the dynamical core of the \texttt{MITgcm} \citep{adcroft_implementation_2004} to the radiative transfer solver of \texttt{petitRADTRANS} \citep{molliere_petitradtrans_2019}. The model includes Rayleigh friction at the top (at $10^{-5}$~bar) and the bottom (at $700$~bar) of the computational domain in order to stabilize the atmosphere against unphysical gravity waves and to mimic Ohmic dissipation \citep{carone_equatorial_2020}. 

    We use a stellar effective temperature and stellar radius of HATS-6 of 3724~K and $0.57~R_\odot$, respectively, and assume HATS-6b to be tidally locked on a circular orbit with an orbital separation and period of $0.03623$~AU and $3.3252725$~days, respectively. \refree{The surface gravity is assumed to be $7.94~\mathrm{ms}^{-2}$ and the internal temperature is set to $T_\mathrm{int} = 50$~K following \citet{thorngren_intrinsic_2019}.}
    
    The thermal forcing part of \texttt{expeRT/MITgcm} solves the radiative transfer equation including isotropic scattering using the Feautrier method \citep{feautrier_sur_1964} with approximate lambda iteration \citep{olson_rapidly_1986}.  \texttt{expeRT/MITgcm} uses 11 correlated-k wavelength bins for the radiative transfer calculations \citep[S1 resolution;][]{schneider_exploring_2022}. We chose to use a radiative time-step of $100$~s, which is four times the dynamical time-step\footnote{This is a typical choice for hot Jupiters \citep[see, e.g.][]{showman_atmospheric_2009, lee_simulating_2021, schneider_exploring_2022, schneider_no_2022}}. \texttt{GGchem} \citep{woitke_equilibrium_2018} is used to pre-calculate a grid of chemical equilibrium abundances that are then used to generate a premixed opacity grid for \texttt{expeRT/MITgcm}. The elemental abundances are assumed solar \citep{asplund_chemical_2009} with a solar C/O ratio of C/O = 0.54 and scaled for the metalicity of HATS-6b of [Fe/H] = $0.2 \pm 0.09$ \citep{hartman_hats-6b_2015}. A gas temperature $T_{\rm gas}$ and gas pressure $p_{\rm gas}$ grid for 100 K - 10000 K, 10$^{-5}$ bar - 650 bar is calculated. Taking the temperature and pressure averages from this premixed grid, the specific gas constant $R = 3925~\mathrm{J}~\mathrm{kg}^{-1}\mathrm{K}^{-1}$ and specific heat capacity at constant pressure of $c_p = 14966~\mathrm{J}~\mathrm{K}^{-1}$ are obtained. 
    
    Most absorption cross sections are obtained from \texttt{exomol}\footnote{\url{https://www.exomol.com/}}. H$_2$O, CO$_2$, CH$_4$, NH$_3$, CO, H$_2$S, HCN, PH$_3$, TiO, VO, FeH, Na, and K opacities are used for the gas absorbers\footnote{References can be found in Table.~\ref{tab:Appendix_crosssecs}.}. Furthermore, we include Rayleigh scattering with H$_2$ \citep{dalgarno_rayleigh_1962} and He \citep{chan_refractive_1965} and collision-induced absorption (CIA) with H$_2$--H$_2$ and H$_2$--He \citep{borysow_collision-induced_1988, borysow_collision-induced_1989-1, borysow_collision-induced_1989, borysow_high-temperature_2001, richard_new_2012, borysow_collision-induced_2002} and H$^-$ \citep{gray_observation_2008}.

    \subsection{Cloud structure}
    \label{sec:Model_clouds_static}
    
    The results from \texttt{expeRT/MITgcm} are used as input to a kinetic cloud formation model. 1D ($T_{\rm gas}$(z), $p_{\rm gas}$(z), $v_{\rm z}(z)$)-profiles are extracted for each point in a longitude-latitude grid, with spacing of $45^\circ$ in longitude ($\phi = \{ -135^\circ, -90^\circ, -45^\circ, 0^\circ, 45^\circ, 90^\circ, 135^\circ, 180^\circ \}$) and $\sim22.5^\circ$ in latitude ($\theta = \{ 0^\circ, 23^\circ, 45^\circ, 68^\circ, 86^\circ \}$), similar to previous works \citep{helling_mineral_2016, helling_understanding_2019, helling_sparkling_2019, helling_mineral_2020, helling_cloud_2021, samra_clouds_2022, carone_wasp-39b_2023, helling_exoplanet_2023}. $T_{\rm gas}$(z) [K] is the local gas temperature, $p_{\rm gas}$(z) [bar] is the local gas pressure, and $v_{\rm z}$(z) [cm s$^{-1}$] is the local vertical velocity component of the 1D profile extracted from the 3D GCM. Each 1D model was run top-down in the atmosphere until it reached at least 0.1~bar. For the southern hemisphere ($\theta < 0^\circ$), all grid points are mirrored across the equator ($\theta = 0^\circ$). \texttt{GGChem} is used to determine the local gas-phase composition in chemical equilibrium for which solar elemental abundances from \citet{asplund_chemical_2009} are initially assumed which are scaled for the metallicity of HATS-6b.

    \subsubsection{Cloud formation}
    
    Our kinetic cloud formation model treats the micro-physical processes of nucleation, bulk growth, and evaporation in combination with gravitational settling, element consumption, and replenishment using the moment method for the moments $L_{\rm j}(V)$ in volume space $V$ \citep{gail_primary_1986, gail_dust_1988, dominik_dust_1993, woitke_dust_2003, woitke_dust_2004, helling_dust_2004, helling_dust_2006, helling_dust_2008}. Cloud condensation nucleii (CCNs) are considered to form from four nucleating species: TiO$_2$, SiO, KCl, and NaCl. This choice was based on the ones included in previous studies \citep{lee_dust_2015, bromley_under_2016, lee_dust_2018, gao_microphysics_2018, kohn_dust_2021, sindel_revisiting_2022, kiefer_effect_2023}. However, we note that there are current efforts to expand the list of nucleating species \citep[e.g.][]{gobrecht_bottom-up_2022, gobrecht_bottom-up_2023, lecoq-molinos_vanadium_2024}. The growth of the mixed material cloud particles is modelled through the surface growth of 16 bulk materials: \ce{TiO2}[s], \ce{Mg2SiO4}[s], \ce{MgSiO3}[s], MgO[s], SiO[s], \ce{SiO2}[s], Fe[s], FeO[s], FeS[s], \ce{Fe2O3}[s], \ce{Fe2SiO4}[s], \ce{Al2O3}[s], \ce{CaTiO3}[s], \ce{CaSiO3}[s], KCl[s], \ce{NaCl[s]}. These materials grow and evaporate through 132 surface reactions \citep{helling_sparkling_2019} onto the surface of the CCN. The formation and evaporation of cloud particles depletes and replenishes the 11 participating elements Mg, Si, Ti, O, Fe, Al, Ca, S, Na, K and Cl, which affects the gas-phase equilibrium abundances. The kinetic cloud model assumes a cloud free element reservoir deep in the planet. Through mixing, the upper layers are replenished with cloud forming elements. The exact strength of mixing within exoplanet atmospheres is difficult to determine \citep{parmentier_3d_2013, steinrueck_effect_2019, samra_clouds_2022}. For this study, we used a quasi-diffusive approach utilising the standard deviation of neighbouring cells to compute a relaxation timescale (see Appendix B in \citet{helling_exoplanet_2023}). Previous work \citep[e.g.][]{gao_aerosol_2020} considered sulfur, manganese and zinc-bearing species which typically condense below 1000 K for pressure ranges between $10^{-4}$ bar to $10^{2}$ bar. However, these elements are much less abundant and therefore do not contribute significantly to the cloud particle material composition.
    
    The resulting cloud particle material volume fractions $V_{\rm s}/V_{\rm tot}$, where $V_{\rm s}$ is the cloud particle volume of a given species s and $V_{\rm tot}$ the total cloud particle volume, average particle size $\langle a \rangle$ [cm], and cloud particle number density $n_{\rm d}$ [cm$^{-3}$] are provided as input for the opacity calculation in the 3D GCM. All cloud particles are assumed to be spherical.

    \subsubsection{Cloud formation mixing treatment}

    \refree{Since our cloud model calculates the cloud structure in the stationary case, the elemental replenishment has to balance the depletion through nucleation, bulk growth, and gravitational settling \citep{woitke_dust_2004}. For each atmospheric layer, the rate at which elements are replenished from the deep atmosphere is described by the mixing timescale $\tau_\mathrm{mix}$. This timescale is calculated using the vertical velocities extracted from the GCM as described in Appendix~B in \cite{helling_exoplanet_2023}. Consistent with our cloud formation model, the mixing timescale is therefore derived locally and is not constant through the atmosphere \citep{helling_understanding_2019, helling_cloud_2021}. Replenishment is also sometimes described as a diffusive process which is common for chemical kinetics models where the diffusion coefficient is called $K_\mathrm{zz}$ \citep[e.g.][]{moses_disequilibrium_2011, agundez_pseudo_2014, tsai_vulcan_2017, baeyens_grid_2021, konings_impact_2022}. Other studies have used $K_\mathrm{zz}$ to derive a replenishment timescale through $\tau_\mathrm{dif} \sim {H_p^2}/{K_\mathrm{zz}}$ where $H_p$ [cm] is the scale height \citep[see e.g.][]{charnay_self-consistent_2018, helling_sparkling_2019}. The difference between $\tau_\mathrm{dif}$ and $\tau_\mathrm{mix}$ is that $\tau_\mathrm{dif}$ describes the exchange between adjacent atmospheric layers whereas $\tau_\mathrm{mix}$ describes the rate at which elements are replenished from the deep atmosphere through convective processes to a given atmospheric layer.}

    A hallmark of the climate of close-in, tidally locked gas-giants is a strong equatorial wind jet \citep{showman_equatorial_2011}. The high hydrodynamic velocities lead to advection timescales that are typically orders of magnitude shorter than the gravitational settling or diffusion of cloud particles \citep{woitke_dust_2003, powell_two-dimensional_2024}. The nucleation and bulk growth timescales of cloud particles on the other hand can still be shorter than the advection timescales \citep{helling_dust_2006, lee_dust_2018,  powell_two-dimensional_2024, kiefer_fully_2024}. Studies which considered the horizontal transport of cloud particles \citep[e.g.][]{lee_dynamic_2016, lines_simulating_2018, komacek_patchy_2022, lee_modelling_2023, powell_two-dimensional_2024} suggest that the number density and size of cloud particles are more longitudinally uniform than studies which only considered vertical mixing as cloud particle transport mechanism \citep[e.g.][]{helling_cloud_2021, roman_clouds_2021, samra_clouds_2022, helling_exoplanet_2023}. \refree{Similarly, the replenishment of gas-phase species can happen through both horizontal and vertical mixing. Which of the two dominates depends on the atmospheric conditions. Studies comparing vertical and horizontal mixing timescales for gas-phase species found that at pressures below 0.1 bar, vertical mixing often dominates \citep{helling_understanding_2019, baeyens_climate_2021, zamyatina_quenching-driven_2024}. In general,} considering the horizontal transport of cloud particles is computationally expensive. Studies evaluating the effect of horizontal transport are either limited to few simulation days \citep[e.g.][]{lee_dynamic_2016, lines_simulating_2018}, make simplifying assumptions on the cloud formation \citep[e.g.][]{ komacek_patchy_2022, roman_clouds_2021, lee_modelling_2023}, or make simplifying assumptions on the horizontal transport \citep[e.g.][]{powell_two-dimensional_2024}.

    \subsubsection{Numerical aspects of cloud-GCM interface}

    The 3D GCM employs a sponge layer to stabilize the upper boundary and basal Rayleigh drag to stabilize the lower boundary. While both layers ensure numerical stability, they are also physically justified.
    
    The cloud model that interfaces with the GCM, similarly, requires numerical stabilisation measures that are justified by physical mechanisms that limit the condensate cloud model. At the upper boundary of the modelling domain,  the growth of the cloud particles is limited by decreasing collisional rates in the upper atmosphere. At the lower boundary, cloud growth is limited by evaporation of cloud materials in the dense and hot deeper atmosphere. To ensure numerical stability, the cloud particle opacities were decreased linearly at the upper and lower pressure limit of the \texttt{expeRT/MITgcm} pressure grid until they reach 0. This decrease prevents a sudden drop in opacities that would otherwise trigger instabilities in the radiative transfer. At the top of the atmosphere, the cloud structure is only interpolated in the upper most grid cell at p = 10$^{-5}$~bar. At the high pressure limit of the cloud structures, the interpolation starts from the the lowest pressure of the cloud structure which is at pressures higher than 10$^{-1}$~bar.
    
    The \texttt{expeRT/MITgcm} uses a cubed-sphere C32 grid \citep{adcroft_implementation_2004}. The cells of this grid are more uniformly distributed than a longitude-latitude grid which prevents overcrowding at the poles. The resolution of each cell in the C32 grid is approximately 2.8 by 2.8 degrees. The interpolation from the low resolution longitude-latitude cloud model grids to the \texttt{expeRT/MITgcm} cubed-sphere grid was done using two interpolation steps to reduce interpolation artifacts while keeping the structure of the low resolution grid. First, a bilinear interpolation to a longitude-latitude grid with cell size $\Delta$lon = $\Delta$lat = 3 degrees was performed. Afterwards, a bilinear interpolation to a cubed-sphere grid is used. During runtime, clouds are added incrementally to prevent sharp changes in the opacity structure. Out of the 2000 total simulation days, the first 100 simulation days are run without clouds. Then, cloud opacities are linearly increased over the next 100 simulation days until they are fully added at simulation day 200.
    
    The step-wise iterative process applied here is a hands-on version of the iteration processes executed in every complex model implemented as, for example, an operator splitting method in \cite{helling_dust_2001}. Such methods make use of the time-scale difference that may occur, for example, between condensation and hydrodynamical processes. In the case of cloud formation, the formation processes modelled here (nucleation, grow and evaporation) act on very short time scales since they predominantly occur in collisionally dominated gases in exoplanet atmospheres. This may, however, change in the upper atmosphere regions between $10^{-4}-10^{-5}$~bar where the densities are so low that e.g. photochemistry affects the gas-phase. In this work, photochemistry is taken into account in post-processing. The majority of the cloud formation occurs at deeper levels and, generally, cloud haze models that extend higher up \citep{steinrueck_3d_2021} still find that the hydrodynamic assumption for these layers is adequate. Further, we aim to resolve here the atmosphere regions accessible in the infrared by JWST which are typically at pressures higher than $10^{-4}$~bar.
    
    Possible long-term changes of the thermodynamic atmosphere structure may be linked to the deep atmosphere which can take more than 80000 simulation days to converge \citep{wang_extremely_2020, schneider_no_2022}. Similarly, 1D time dependent cloud models have shown that convergence times can reach up to 8000 simulation days \citep{woitke_dust_2020}. Fully coupled cloudy GCMs are computationally intensive and therefore often limited to evaluation times below 5000 simulation days \citep[e.g.][]{lee_dynamic_2016, lee_dynamic_2017, roman_modeling_2017, roman_modeled_2019, lines_exonephology_2018, lines_simulating_2018, roman_clouds_2021, komacek_patchy_2022}. However, recent run-time improvements of GCMs \citep{schneider_exploring_2022} might allow for longer simulation times in the future. It should be noted, however, that this work showed only a minor change in the upper atmosphere structure ($p<1$~bar) after 2000~days simulation time (see Appendix \ref{sec:App_gcm_convergence}).

    \subsubsection{Cloud opacities}
    \label{sec:Model_clouds_opacities}

    To calculate the interaction between the photons and the cloud particle, we use Mie-theory \citep{mie_beitrage_1908}. This theory is an analytical solution to the Maxwell equations under the assumption of spherical particle with an effective refractive index $\epsilon_\mathrm{eff}$. 
    
    To find the effective refractive index $\epsilon_\mathrm{eff}$ of a given mixture of bulk material, we follow the approach of \citet{lee_dynamic_2016} and start by using the Bruggeman mixing rule \citep{bruggeman_berechnung_1935}. In case of non-convergence, we fall back to the Landau-Lifshitz-Looyenga method \citep{looyenga_dielectric_1965}. The homogeneous opacity values for all species $s$ used in this paper can be found in the Appendix \ref{sec:App_opadata}. Using the effective refractive index, Mie-theory is used to calculate the absorption efficiency $Q_\mathrm{abs}$, the scattering efficiency $Q_\mathrm{sca}$ and the anisotropy factor $g$. From these, the wavelength dependent absorption coefficient $\kappa_\mathrm{abs}^\mathrm{cloud} (\lambda)$ [cm$^2$ kg$^{-1}$] and the scattering coefficient $\kappa_\mathrm{sca}^\mathrm{cloud} (\lambda)$ [cm$^2$ kg$^{-1}$] are calculated:
    \begin{align}
        \kappa_\mathrm{abs}^\mathrm{cloud}(\lambda) &= \pi \langle a \rangle^2 \frac{n_\mathrm{d}}{\rho_\mathrm{gas}}  ~Q_\mathrm{abs} (\langle a \rangle, \lambda, \epsilon_\mathrm{eff}) \\
        \kappa_\mathrm{sca}^\mathrm{cloud}(\lambda) &= \pi \langle a \rangle^2 \frac{n_\mathrm{d}}{\rho_\mathrm{gas}} ~Q_\mathrm{sca} (\langle a \rangle, \lambda, \epsilon_\mathrm{eff}) ~ (1 - g) 
    \end{align}
    where $n_\mathrm{d}$ [cm$^{-3}$] is the number density of cloud particles, $\langle a \rangle$ [cm] the mean cloud particle radius, $\rho_\mathrm{gas}$ [g cm$^{-3}$] the gas density, and $\lambda$ [cm] the wavelength of the photon. 
    Our cloud model uses the moment method which allows the efficient modelling of heterogeneous cloud particles without an explicit cloud particle size distribution. Reconstructing the full size distribution of cloud particles \citep{helling_dust_2008} would require to assume a functional form. We therefore assume monodisperse cloud particles with a local mean particle size derived from the kinetic cloud model as used in \citet{helling_mineral_2020}. A study on the effect of the assumptions of monodisperse cloud particles can be found in \citet{samra_mineral_2020}. The absorption and scattering coefficients are then added to the gas-phase opacities of the radiative transfer of \texttt{expeRT/MITgcm}:
    \begin{align}
        \kappa_\mathrm{abs}^\mathrm{tot}(\lambda) &= \kappa_\mathrm{abs}^\mathrm{gas}(\lambda) + \kappa_\mathrm{abs}^\mathrm{cloud}(\lambda) \\
        \kappa_\mathrm{sca}^\mathrm{tot}(\lambda) &= \kappa_\mathrm{sca}^\mathrm{gas}(\lambda) + \kappa_\mathrm{sca}^\mathrm{cloud}(\lambda)
    \end{align}
    where $\kappa_\mathrm{abs}^\mathrm{gas}$ [m$^2$ kg$^{-1}$] and $\kappa_\mathrm{sca}^\mathrm{gas}$ [m$^2$ kg$^{-1}$] are the absorption and scattering coefficients for the gas, respectively. $\kappa_\mathrm{abs}^\mathrm{tot}$ [m$^2$ kg$^{-1}$] and $\kappa_\mathrm{sca}^\mathrm{tot}$ [m$^2$ kg$^{-1}$] are the total absorption and scattering coefficients, respectively.

    \subsection{Stopping the step-wise iteration}
    \label{sec:Model_stopping}
    One of the aims of this work is to demonstrate that a 3D atmosphere solution including detailed cloud formation is computationally feasible for warm Saturn type exoplanets. We do this through a step-wise iteration between the two modelling complexes. This enables the full complexity of both the 3D GCM and the cloud model, to achieve a conceivable accuracy for a global exoplanet atmosphere structure.
    
    Here, the conceivable accuracy is determined from the observational accuracy, hence, the stopping criterion depends on the observational facilities used. For this project, the spectral precision of transmission spectra from the JWST instruments NIRspec Prism and MIRI LRS after two and after ten transit measurements are used as the primary stopping criterion. After each iteration, a transmission spectrum is calculated (Sect.~\ref{sec:Model_clouds_trans}) and compared to the previous iteration step. If the observational differences between iterations fall below the spectral precision (dotted line in Fig.~\ref{fig:gcm_residuals}), the step-wise iteration is stopped. Since our model iterates between the cloud model and the GCM, the main goal is to have no observable impact of the changing cloud structure on the transmission spectra. The abundances in chemical disequilibrium are calculated in post-processing and the impact on the transmission spectra is used as a secondary stopping criteria.

    Additional further stopping criteria may be derived from the ($T_{\rm gas}$, $p_{\rm gas}$) - structures and the cloud opacity. As will be demonstrated later, once changes in transmission spectra fall below the observable accuracy, the cloud properties between the two successive iterations remain similar. This suggests that the still existing changes between the ($T_{\rm gas}$, $p_{\rm gas}$)-structures, which may be substantial but locally confined, do not affect the gas-phase and cloud opacity structure sufficiently enough to change the spectrum beyond spectral precision. Hence, these changes will not affect the interpretation of the data.

    \subsection{Disequilibrium gas-phase chemistry}
    \label{sec:Model_argo}
     The disequilibrium chemistry for the H/C/N/O complex of HATS-6b is modelled to assess how each iteration affects the atmospheric chemistry, and as a result the transmission spectra. This is done using an updated version of the chemical network STAND2020 \citep{rimmer_chemical_2016, rimmer_erratum_2019, rimmer_hydrogen_2019} in combination with the 1D photochemistry and diffusion code \texttt{ARGO} \citep{rimmer_chemical_2016}. \texttt{ARGO} models chemical transport, wavelength-dependent photochemistry, and cosmic ray transport by following a parcel of gas as eddy diffusion leads it up and down through the atmosphere (further description in e.g. \citet{rimmer_chemical_2016} and \citet{barth_moves_2021}). STAND2020 is a chemical H/C/N/O network containing the reaction rates for more than 6600 reactions in the temperature range of 100 K to 30000 K. STAND2020 involves all reactions for species containing up to six H, two C, two N, and three O atoms and also contains reactions with He, Na, Mg, Si, Cl, Ar, K, Ti, and Fe bearing species. The network has been tested for the atmospheres of Earth and Jupiter \citep{rimmer_chemical_2016} in addition to a number of hot-Jupiter models \citep{barth_moves_2021, hobbs_sulfur_2021, tsai_photochemically_2023}. The chemical network is run through the 1D photochemistry and diffusion code, \texttt{ARGO}, which models chemical transport and the effect of photochemistry and cosmic rays.

    The inputs for \texttt{ARGO} and STAND2020 have been chosen as follows:
    
    \begin{enumerate}
        \item ($T_{\rm gas}$, $p_{\rm gas}$) profiles and vertical eddy diffusion profile: Eight different 1D profiles are extracted from the output of the \texttt{expeRT/MITgcm} by averaging over areas of the 3D grid. The eight profiles are six terminator regions with the longitudes ($\phi = \{90^\circ,270^\circ \}$) and latitudes ($\theta = \{ 0^\circ, 23^\circ, 68^\circ\}$), and the sub-stellar and anti-stellar points. 
        \item Atmospheric element abundances: Solar abundances adapted for metallicity ([Fe/H] = 0.2) in accordance with the initial abundances used for the GCM.
        \item Stellar XUV spectrum driving photochemistry: Spectrum obtained from the MUSCLES survey where the M1.5V star, GJ667C, is chosen as a proxy to HATS6. The spectrum covers the XUV range and is composed of a combination of modelled and observed spectra \citep{france_muscles_2016, youngblood_muscles_2016, youngblood_muscles_2017, loyd_muscles_2016, loyd_muscles_2018}. The XUV spectrum used in this study can be seen in Appendix \ref{fig:app_sed}.
        \item Cosmic rays: Implemented based on the ionization rate of low energy cosmic rays (LECR) as explained by \citet{rimmer_ionization_2013} and \citet{barth_moves_2021}.
        \item Stellar energetic particles: To account for the higher activity of M-dwarf host stars, SEPs have been included in the model. The SEPs are introduced by scaling the spectrum of a solar SEP event to fit the host star based on the X-ray flare intensity (method further described in \citet{barth_moves_2021}). A number of X-ray flare intensities has been reported for M-dwarf stars \citep{hunsch_x-ray_2003, namekata_optical_2020} ranging from $\sim$ 0.001 to 0.2 W m$^{-2}$ at 1 AU.
        For this study we chose to implement SEPs corresponding to X-ray flare intensities of 0.1 W m$^{-2}$ at 1AU. The effect of the SEPs will be included continuously throughout the run, and not as a finite event.
    \end{enumerate}

    Different locations on the planet will experience a different influx of stellar radiation and SEPs, and the model inputs are therefore varied accordingly. For the sub-stellar point, both the stellar spectrum and SEPs are included. For the anti-stellar point, neither the stellar spectrum nor SEPs are included. For the terminator coordinates, only SEPs are included. The reason for including SEPs but not the stellar spectrum for the terminator region is that the shallow angle of incidence for radiation from the host star causes the radiation to pass through so much atmospheric layers before it reach the bulk of the 1D simulated atmosphere profile above the terminator that its influence is negligible. Since XUV radiation is easily scattered by the atmosphere, the stellar spectrum is not included for the terminator regions, whereas SEPs have been shown to penetrate deeper into the atmosphere \citep{barth_moves_2021} and are therefore included. The SEPs are not scaled based on the incident angle. The output of the STAND2020-\texttt{ARGO} run is the relative concentrations (n$_i$/n$_{gas}$) of more than 511 gas-phase species. Based on these relative concentrations as well as the cloud opacities mentioned previously, the transmission spectra can be produced.

    \subsection{Transmission radiative transfer}
    \label{sec:Model_clouds_trans}
    
    The transmission spectrum of HATS-6b is produced by adding the cloud opacities to \texttt{petitRADTRANS} \citep{molliere_petitradtrans_2019, molliere_retrieving_2020, alei_large_2022}, where the cloud opacities from the micro-pyhsical cloud model are included as opacity source (Sect.~\ref{sec:Model_clouds_opacities}). Each transmission spectrum uses the input cloud structure for the GCM, the temperature structure of the GCM, and the post-processed gas-phase relative concentrations from ARGO. The transmission spectrum of the cloudless GCM run (iteration 0) is calculated cloud free.
    
    To calculate the transmission spectrum, the terminator region is divided into ten equally spaced, and equally sized grid cells. Each cell covers a latitudinal range of $\Delta$lat = 22.5$^\circ$. Choosing a constant longitudinal range would lead to the neglect of data points within the terminator region, especially close to the poles. Therefore, a latitude dependent longitudinal range was chosen to ensure equally sized grid cells. At the equator the longitudinal range is $\pm\Delta$lon = 10$^\circ$ \citep{lacy_one_jwst_2020}. The total transmission spectrum is then calculated as the average of the transmission spectra from all regions:
    \begin{equation}
        R_\mathrm{tot} (\lambda) = \sqrt{\frac{1}{N}\sum_{i=1}^{N} R_i^2(\lambda)}
    \end{equation}
    Similar to previous studies \citep{lee_exoplanetary_2019, baeyens_grid_2021, nixon_aura-3d_2022}, the following gas-phase species were considered as line opacity species\footnote{References can be found in Table.~\ref{tab:Appendix_crosssecs}.}: H$_2$O, CO, CH$_4$, CO$_2$, C$_2$H$_2$, OH, NH$_3$, and HCN. Furthermore, H$_2$ and He were considered as Rayleigh scattering opacities, and collision-induced absorption (CIA) from H$_2$--H$_2$ and H$_2$--He.

\section{Evaluation of the combined model}
\label{sec:res}

    To study the atmosphere of the warm-Saturn HATS-6b, 6 iterations using \texttt{expeRT/MITgcm} and the kinetic cloud model have been conducted. After each iteration of the GCM, the thermodynamic structure is used to produce a cloud-structure which is included in the next iteration of the GCM. The cloud-free simulations are shown in Sect.~\ref{sec:res_cloudless}. The effect of clouds are presented in Sect.~\ref{sec:res_impact}. After each GCM run, the disequilibrium chemistry (Sect.~\ref{sec:sim_chem}) is post-processed and the differences in the transmission spectrum are determined (Sect.~\ref{sec:sim_diff}).

    \begin{figure*}
        \centering
        \includegraphics[trim={3cm 0cm 3cm 0cm},width=\hsize]{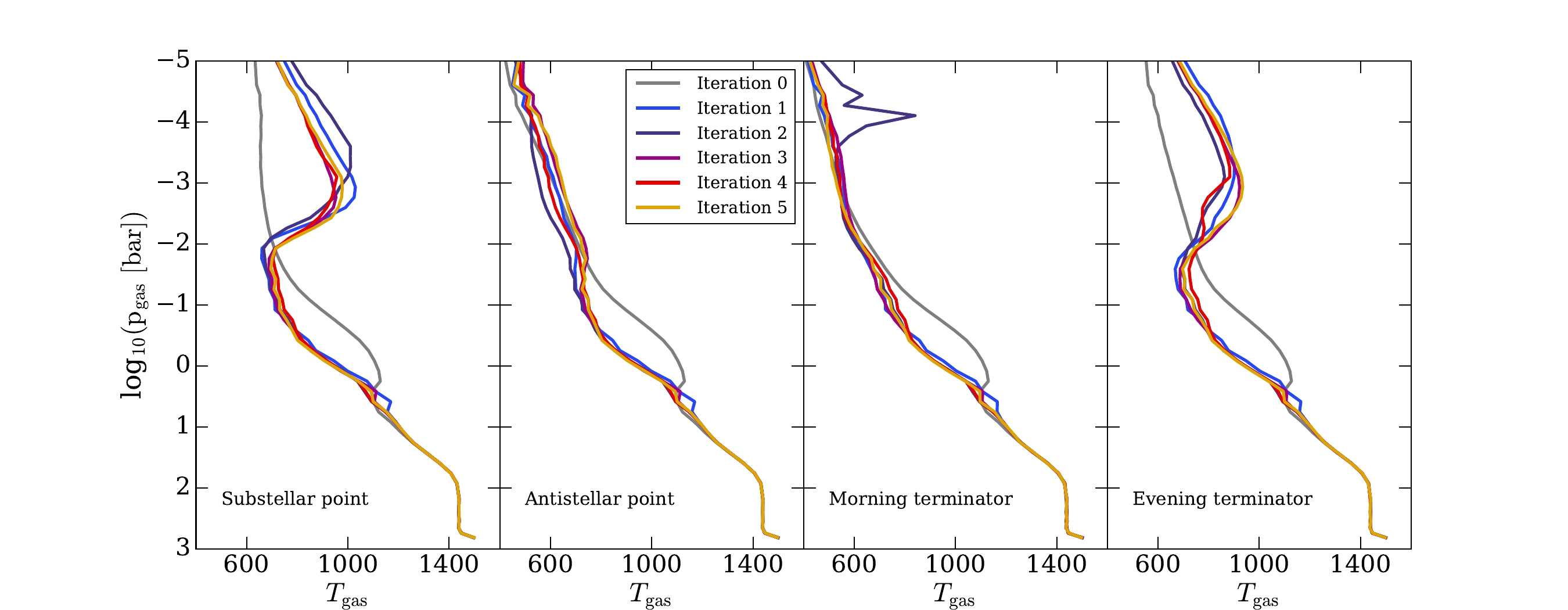}
        \caption{($T_{\rm gas}$, $p_{\rm gas}$)-profiles of the sub-stellar point, anti-stellar point, evening terminator, and morning terminator.}
        \label{fig:gcm_tp}
    \end{figure*}
    
    \begin{figure*}
        \centering
        \includegraphics[trim={3cm 0cm 3cm 0cm},width=\hsize]{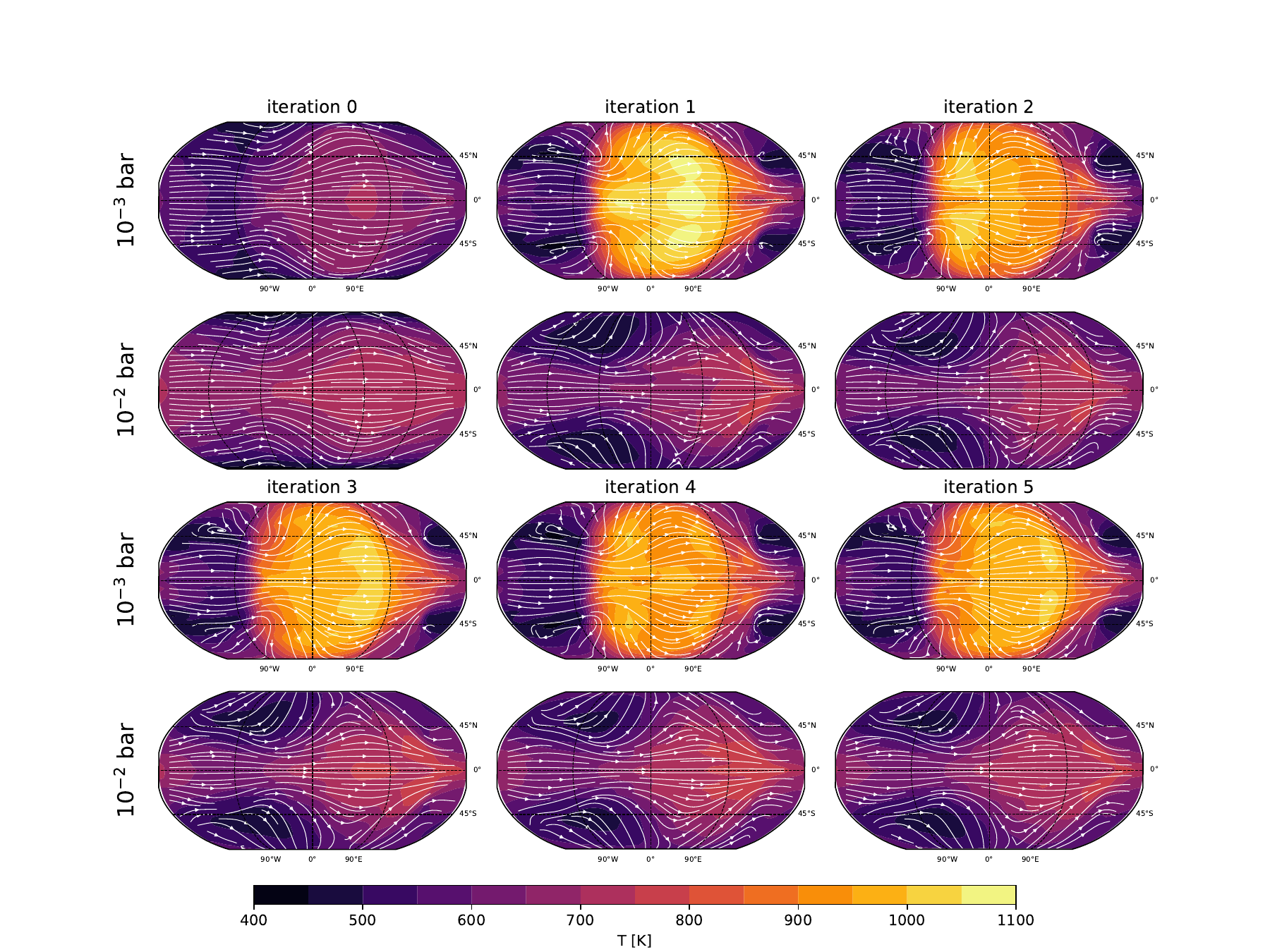}
        \caption{Isobaric slices of the \texttt{expeRT/MITgcm} runs at t = 2000 simulation days. The white lines indicate the horizontal wind velocity fields.}
        \label{fig:gcm_slices}
    \end{figure*}
    
    \begin{figure*}
        \centering
        \includegraphics[width=\hsize]{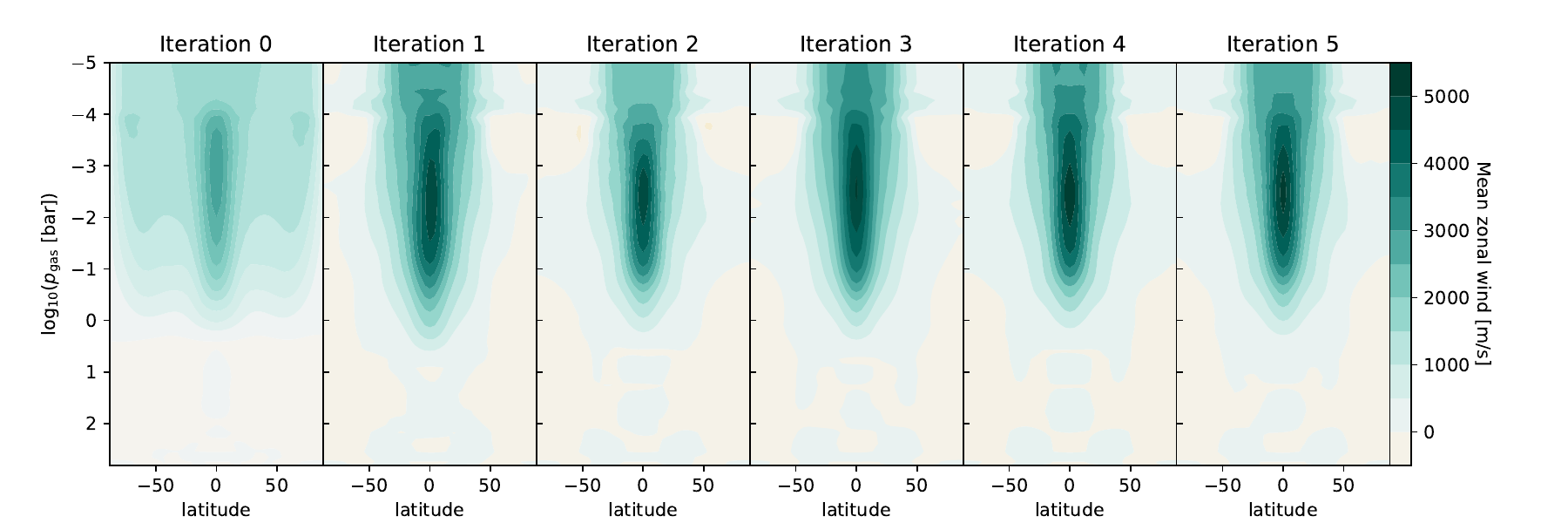}
        \caption{Zonally averaged zonal wind velocities at t = 2000 simulation days.}
        \label{fig:gcm_zonalmean}
    \end{figure*}

    \begin{figure*}
        \centering
        \includegraphics[trim={0cm 0cm 0cm 0cm}, clip, width=0.92\hsize]{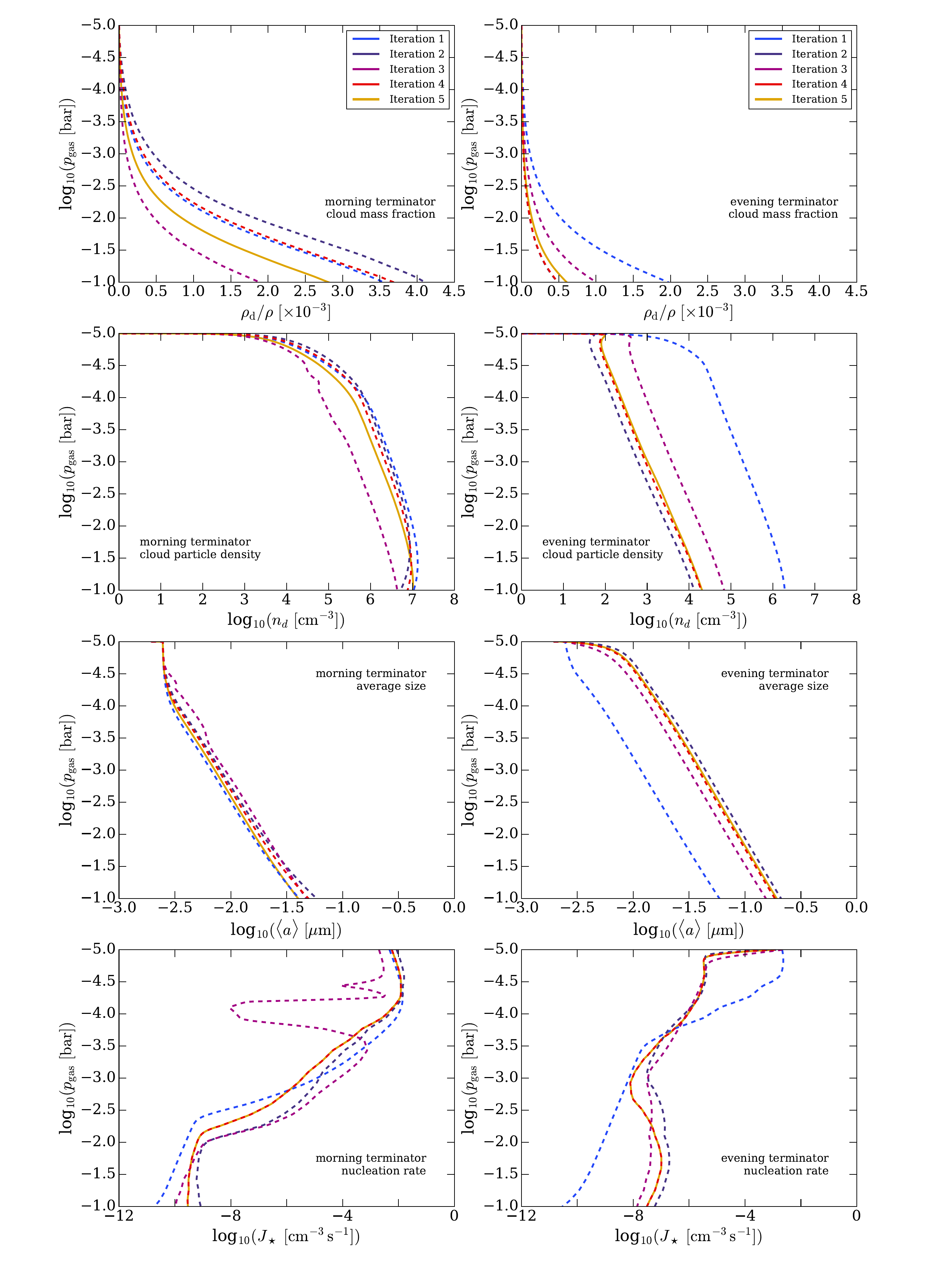}
        \caption{Cloud particle properties of the step-wise iterated cloud structure for the warm Saturn example HATS-6b. Iteration 5 is shown as a solid line to highlight the final result. \textbf{Left:} morning terminator. \textbf{Right:} evening terminator. \textbf{Top:} cloud mass fraction $\rho_\mathrm{d}/\rho$. \textbf{Upper middle:} cloud particle number density $n_d$. \textbf{Lower middle:} average cloud particle size $\langle a \rangle$. \textbf{Bottom:} Nucleation rate $J_\star$. The sub- and anti-stellar point can be found in Fig~\ref{fig:cloud_iterations_stellar}.}
        \label{fig:cloud_iterations}
    \end{figure*}

    \begin{figure}
        \centering
        \includegraphics[width=\hsize]{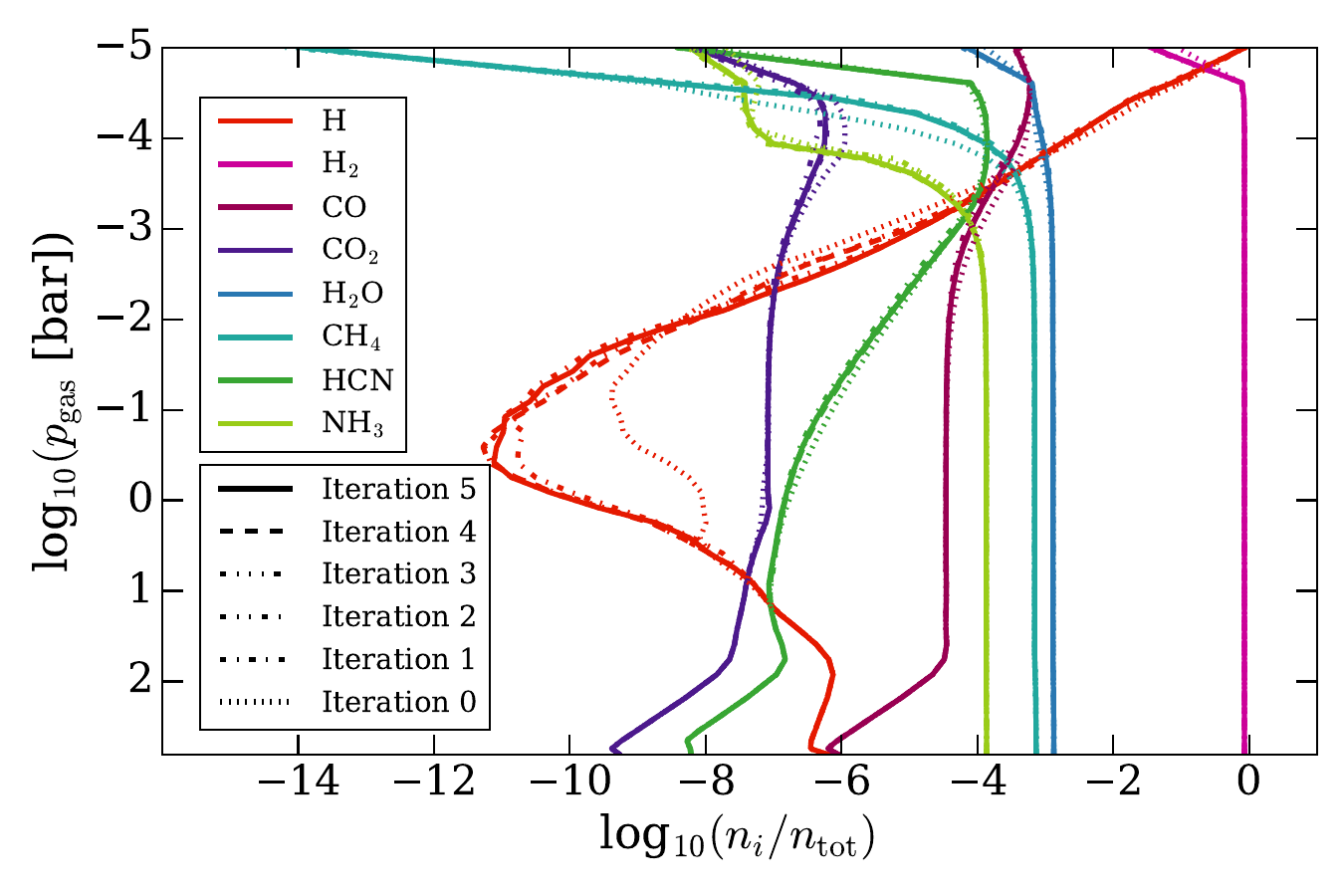}
        \includegraphics[width=\hsize]{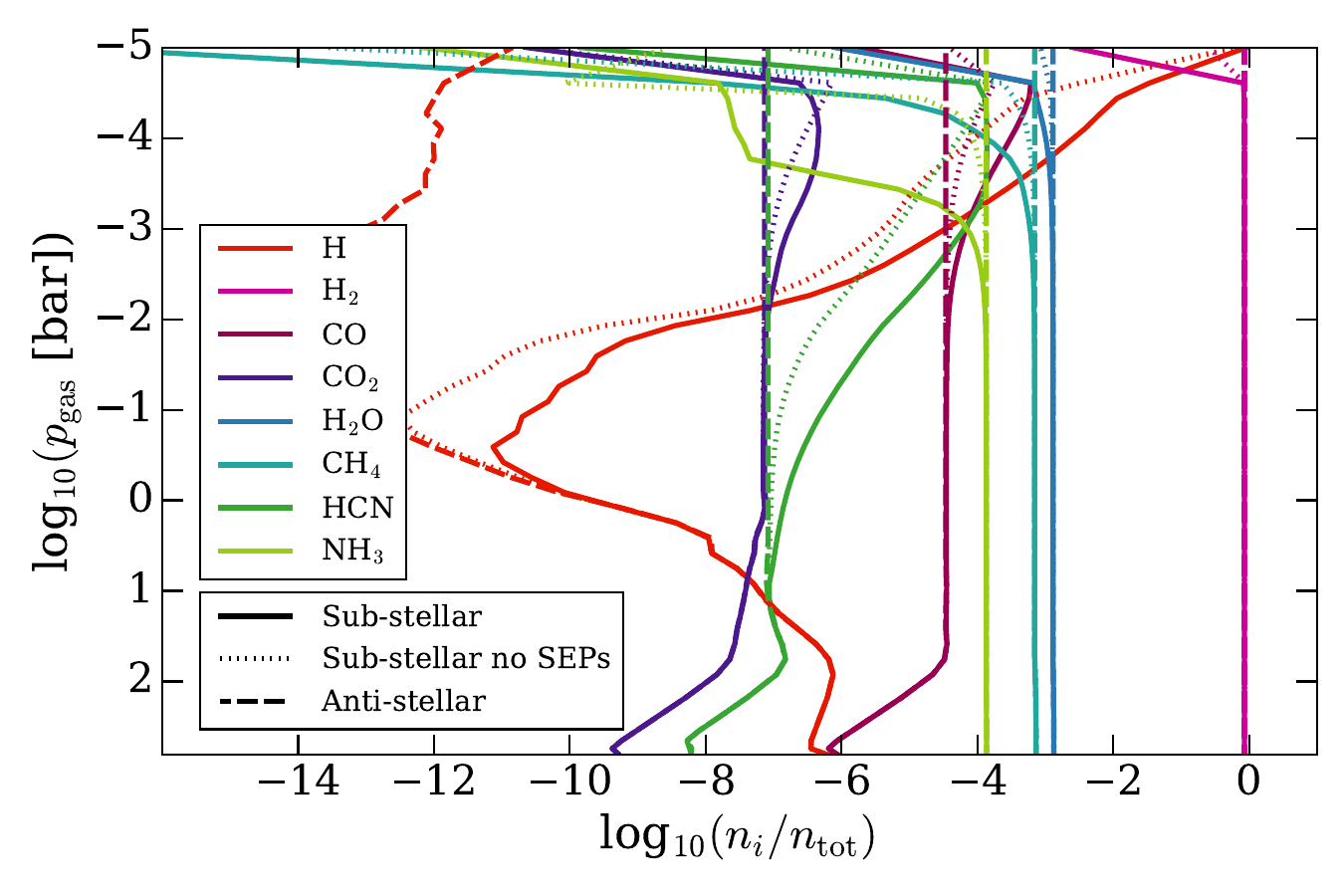}
        \includegraphics[width=\hsize]{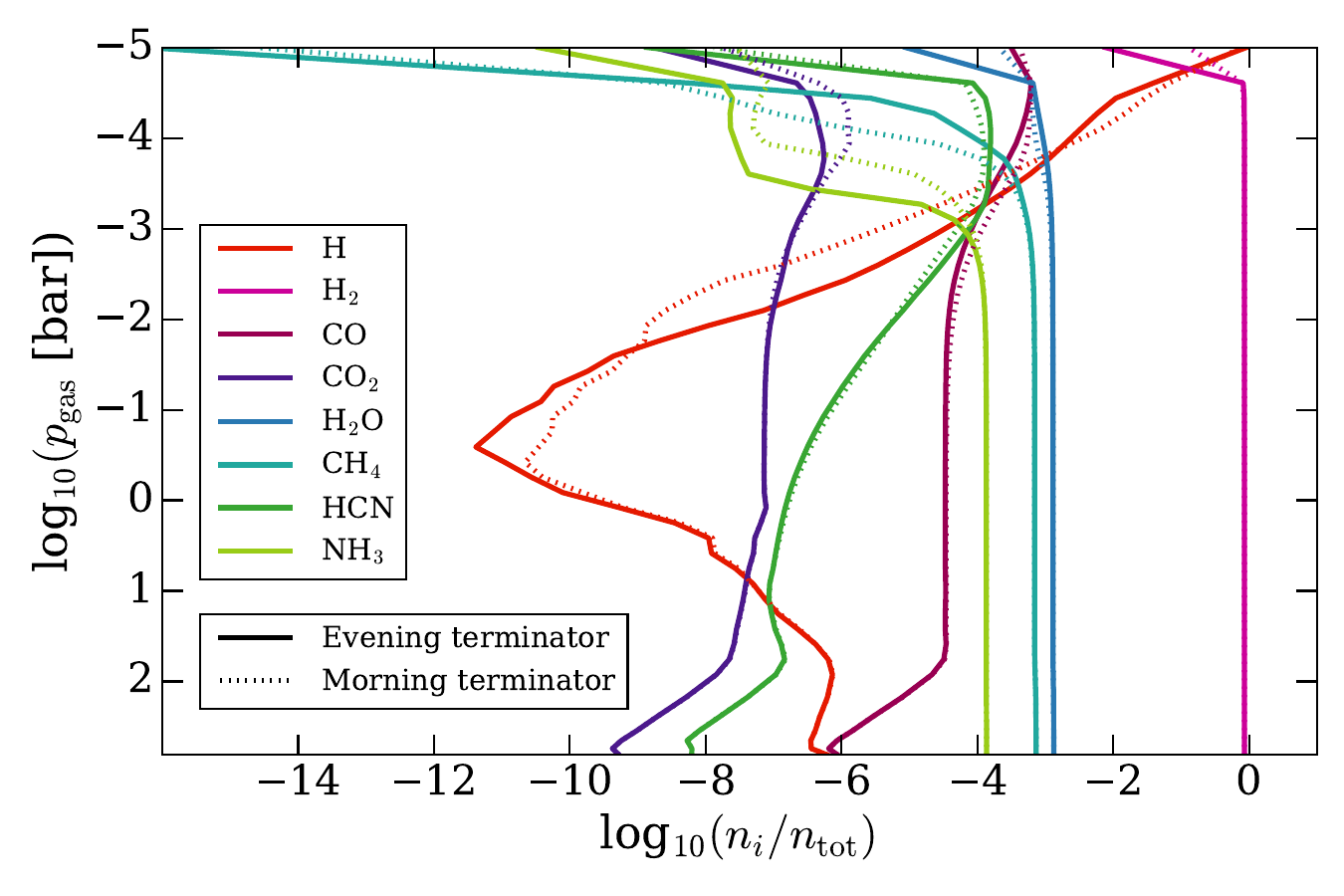}
        \caption{Concentrations of non-equilibrium gas species for the warm Saturn example HATS-6b. \textbf{Top:} Terminator region (both morning and evening) averaged over all six coordinates for all iterations. \textbf{Middle:} Sub- and anti-stellar point for the final iteration (Iteration 5). \textbf{Bottom:} Morning and evening terminators for the final iteration (Iteration 5).}
        \label{fig:abun}
    \end{figure}

    \begin{figure*}
        \includegraphics[width=\hsize]{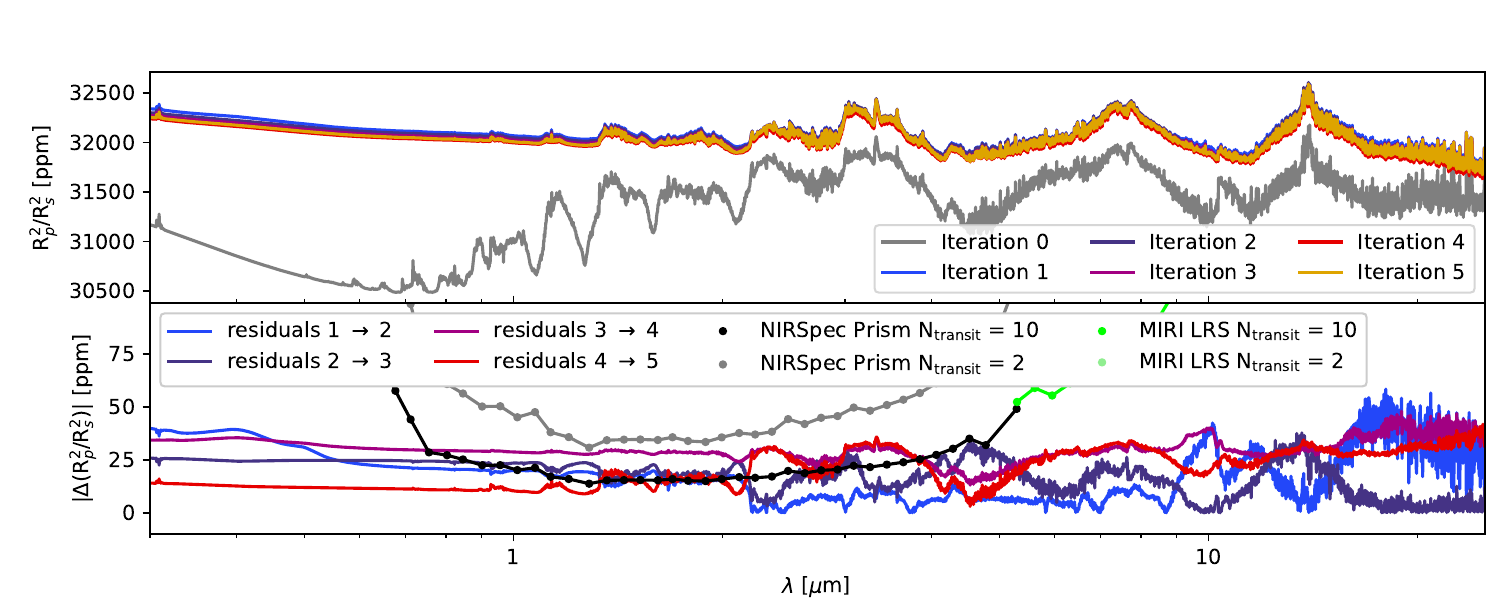}
        \caption{Comparison of the transmission spectra for the warm Saturn example HATS-6b for $\lambda$ = 0.3~$\mu$m to 25~$\mu$m \textbf{Top:} Transmission spectrum for each iteration. \textbf{Bottom:} Absolute residuals between subsequent iterations and the spectral precision for JWST observations with NIRspec Prism and MIRI LRS for 2 and 10 transits.}
        \label{fig:gcm_residuals}
    \end{figure*}

    \subsection{Cloud-free GCM and post-processed clouds}
    \label{sec:res_cloudless}

    The first run, i.e. iteration 0, of the GCM was conducted without clouds. A total of 2000 days were simulated using \texttt{expeRT/MITgcm}. After 2000 days, there was little change in the global average of the ($T_{\rm gas}$, $p_{\rm gas}$)-profiles for pressure layers lower than 1 bar (see Appendix \ref{sec:App_gcm_convergence}). For pressures higher than 1 bar, the temperature did not change by more than 20~K from the initial conditions for the GCM. It is well known, that the convergence deeper in the atmosphere is computationally intensive and difficult to achieve \citep{wang_extremely_2020, skinner_modons_2022, schneider_exploring_2022, schneider_no_2022}. In the cloudy iterations (see Sect.~\ref{sec:res_impact}), the clouds become opaque for pressures higher than $10^{-2}$ bar. Therefore, the atmospheric layers of interest for this study can be assumed to be reasonably converged. 
    
    The ($T_{\rm gas}$, $p_{\rm gas}$) profiles for the sub-stellar point, anti-stellar point, morning terminator, and evening terminator are shown in Fig. \ref{fig:gcm_tp}. The iteration 0 profiles only show minor differences in temperature for pressures greater than 10$^{-1}$ bar. Higher up in the atmosphere, the day-side is roughly 200 K hotter than the night side. The isobaric temperature maps for $p = 1.2 \times 10^{-3}$ bar and $p = 1.2 \times 10^{-2}$ bar can be seen in the left panels of Fig. \ref{fig:gcm_slices}. The zonal mean wind velocities are shown in the left panel of Fig. \ref{fig:gcm_zonalmean} which describe the longitudinal average of winds parallel to the equator and are known to redistribute heat longitudinally \citep{vuitton_composition_2021}. Overall, the global temperature and wind velocity structure of HATS-6b show the characteristics of a highly irradiated gas-giant including a super-rotating equatorial jet stream. This jet stream transports heat from the dayside to the nightside and causes a hot spot offset \citep[e.g.][]{showman_equatorial_2011, cowan_thermal_2012, dang_detection_2018}. In addition to the equatorial jet stream, two weak polar jet streams can be observed.

    Using the output from the cloudless GCM (iteration 0) the cloud structure was post-processed. This cloud structure is labelled iteration 1 because it will be later used as input for the first cloudy iteration of the GCM (iteration 1). The iteration 1 cloud profile of HATS-6b shows global cloud coverage. In Fig.~\ref{fig:cloud_iterations}, the nucleation rate $J_\star$, cloud particle number density $n_d$, average cloud particle size $\langle a \rangle$, and cloud mass fraction\footnote{Also sometimes called dust-to-gas ratio.} $\rho_\mathrm{d}/\rho$ are shown for the morning and evening terminator. The sub-stellar and anti-stellar point can be found in Fig.~\ref{fig:cloud_iterations_stellar}. The cloud particle properties of these profiles are largely the same, although there is an increase in the nucleation rate for the upper atmosphere ($p$ < $10^{-3}$ bar) for the cooler profiles (morning terminator and anti-stellar point) with a corresponding increases in cloud particle number density and cloud mass fraction for these points. The formation of cloud condensation nuclei (CCNs) happens predominantly in the upper atmosphere ($p$ < $10^{-2}$). Though it should be noted that condensation still occurs in these upper regions - hence the increase in average particle size. Going to deeper layers, the cloud particle sizes steadily increase while cloud particle number densities steadily decrease. This inverse correlation between cloud particle number density and cloud particle size is well known \citep[see][]{helling_cloud_2021}. If more cloud particles are present, the same mass of condensation material is spread out over more particles hence reducing the average size of the particles.

    \subsection{Impact of clouds on the atmospheric structure}
    \label{sec:res_impact}

    Iteration 1, 2, 3, 4, and 5 of the GCM are created by including the post-processed cloud opacities from the previous iteration into \texttt{expeRT/MITgcm} as described in Sect.~\ref{sec:Model_clouds_static}. The GCM was again run for 2000 simulation days. Similar to iteration 0, there was little change in the global average temperature above 1 bar after 2000 simulation days (see Appendix \ref{sec:App_gcm_convergence}).

    Going from iteration 0 to iteration 1 highlights the impact of clouds on the temperature and wind structure due to the increased local opacities. The most striking difference is a strong temperature inversion for pressures lower than 10$^{-2}$ bar (higher altitudes) on the sub-stellar point and the evening terminator (Fig. \ref{fig:gcm_tp}). This increased temperature is also visible in the isobaric plots (Fig. \ref{fig:gcm_slices}). At pressures between 10$^{-2}$ bar and 10 bar, the ($T_{\rm gas}$, $p_{\rm gas}$) profiles of iteration 1 show a considerably colder temperature. This drop can be explained with an anti-greenhouse effect (see Sect.~\ref{sec:dis_antig}). For pressures higher than 10 bar, no significant differences in the ($T_{\rm gas}$, $p_{\rm gas}$) profiles were found. Furthermore, iteration 1 has a stronger and narrower equatorial wind jet than iteration 0 (see left panel of Fig. \ref{fig:gcm_zonalmean}). This matches the results of \citet{baeyens_grid_2021}, who showed that for the temperature range of warm Saturns (500 K to 1200 K) an increase in equilibrium temperature results in a faster and narrower jet. The cloud-induced temperature inversion seems to have a similar effect. Furthermore, the weak polar jets of the iteration 0 are not observed in iteration 1. 

    Going from iteration 1 to iteration 2 shows less differences in the thermal and wind structure of HATS-6b than from iteration 0 to 1. However, the temperature profile still changes by up to 90~K. The exception to this is the morning terminator around 10$^{-4}$~bar where a temperature increase of up to 350~K can be seen. This increase in temperature also reduces the nucleation rate which in turn leads to fewer but larger particles (Fig.~\ref{fig:cloud_iterations}). The sudden change in the morning terminator is caused by a dynamical instability caused by the iteration 2 cloud structure which leads to hot air being advected from the day-side into the morning terminator. The general thermal instabilities of the morning terminator are discussed in more detail in Sect.~\ref{sec:dis_dyninst}. While an instability is present in all cloudy iterations, it is more pronounced in iteration 2. The persistence of this instability is a result of the cloud structures being static. Since no other iteration shows a similar behaviour, the temperature increase in the morning terminator of iteration 2 is considered an artefact of the specific configurations of the static clouds. 

    Iteration 3, 4, and 5 all show the same general characteristics as iteration 1 and 2: a temperature inversion around 10$^{-3}$~bar, a cooling around 0.1~bar to 1~bar, a narrow equatorial wind jet, and global cloud coverage. However, the temperature between iterations after iteration 2 still vary by up to $\sim$130~K. Similarly, differences in the temperature structure around 10$^{-3}$~bar can be seen in the isobaric plots (Fig.~\ref{fig:gcm_slices}) and in the zonal mean winds (Fig.~\ref{fig:gcm_zonalmean}). The cloud particle properties on the other hand vary little between iterations 3, 4, and 5. In particular the nucleation rate between iteration 4 and 5 is close to identical. Nevertheless, there are still changes in the cloud particle number density and average size between iteration 4 and 5. As mentioned previously, the goal of this work is to find a solution within the observational accuracy of JWST and therefore only observable changes within the atmospheric structure matter. Whether the changes described here have an observable effect is analysed in Sect.~\ref{sec:sim_diff}.

    \subsection{Disequilibrium chemistry}
    \label{sec:sim_chem}

    To assess the impact of the kinetic gas-phase chemistry and photo-chemistry on the observable atmosphere, and more specifically on transmission spectra of HATS-6b, the disequilibrium chemistry was modelled for each iteration at six coordinates along the terminator region (longitudes ($\phi = \{90^\circ,270^\circ \}$) and latitudes ($\theta = \{ 0^\circ, 23^\circ, 68^\circ\}$)). The resulting gas concentration profiles were averaged over the six coordinates. The relative concentrations $n_i/n_\mathrm{gas}$ of H, H$_2$, CO, CO$_2$, H$_2$O, CH$_4$, HCN, and NH$_3$ can be seen in the top panel of Fig. \ref{fig:abun}. These eight molecules have been chosen due to their high concentrations in the atmospheres or because they are some of the more interesting atmospheric species when looking at the effects of external radiation (e.g. \cite{barth_moves_2021,baeyens_grid_2022}). The concentration profiles show very little variation among the five iterations with clouds, whereas the difference between the cloudless (iteration 0) and cloudy iterations (1 - 5) is somewhat larger. The largest differences between the cloudless and cloudy iterations occurs between $p_{\rm gas} \sim 10^{-2} - 10^{0}$ bar (top of Fig.~\ref{fig:abun}). This pressure range corresponds to a cloud-induced cooling which is present in iterations 1 - 5. Some species (e.g. NH$_3$, CO$_2$, and CH$_4$) also show a difference between cloudy and cloudless at lower pressures (Top panel of Fig.~\ref{fig:abun}) corresponding to a cloud-induced heating in iterations 1 - 5. The cloud-induced heating in the upper atmosphere in combination with a cooling of the lower atmosphere comprises the temperature inversion (see Fig.~\ref{fig:gcm_tp}).
    
    The chemical variations along the equator are illustrated in the middle (sub-stellar and anti-stellar point) and bottom panel (morning and evening terminator). Comparing the top and middle panel we notice that the relative concentration of all species show significantly larger differences between the day- and night-side (the solid and the dashed lines in the middle panel) than the differences between the different iterations (all lines in the top panel). We will further describe the chemistry observed for HATS-6b in Sect. \ref{sec:dis_hats6b}.

    \subsection{Observable differences between iterations}
    \label{sec:sim_diff}

    The goal of the step-wise iteration approach is to reach observational accuracy here chosen to be the JWST NIRspec and MIRI LRS. For HATS-6b, this was reached after six iterations, i.e. after iteration 5. 
    
    All transmission spectra were produced using the temperature structure of the GCM and the chemistry of ARGO. The spectrum of the cloudy iterations included the cloud structures as well. All 6 spectra are shown in the top panel of Fig. \ref{fig:gcm_residuals}. The residuals between iterations 1, 2, 3, 4, and 5 are shown in the bottom panel of the same figure. Additionally, the spectral precision for the JWST instruments NIRSpec Prism \citep{birkmann_near-infrared_2022, ferruit_near-infrared_2022} and MIRI LRS \citep{kendrew_mid-infrared_2015, kendrew_mid-infrared_2016} are shown. The spectral precision was calculated using PandExo \citep{batalha_pandexo_2017} assuming a spectral resolution\footnote{This resolution was chosen to maximise the spectral precision while keeping enough data points to detect gas-phase features.} of R = 10 for two and ten observed transits. If the residuals are significantly larger than the spectral precision, the differences between the iterations are observable. If the residuals are of the same order or lower, observing the differences between iterations will be challenging.

    The transmission spectra of iteration 0 and 1 show clear differences due to the presence of clouds in iteration 1. Compared to iteration 0, the molecular features of iteration 1 are muted for wavelengths below 2~$\mu$m. Around 10~$\mu$m, an additional increase of the relative transit depth can be seen which is caused by silicate species within the cloud particles. Both these effects are known and expected if clouds are present in exoplanet atmospheres \citep{wakeford_transmission_2015, powell_transit_2019, grant_jwst-tst_2023}.

    The residuals in transit depth from iterations 1 to 2 and 2 to 3 are around 25~ppm for wavelengths between 0.7~$\mu$m to 10~$\mu$m. This can be related to a shift in the cloud top, which is defined as the pressure level at which clouds become optically thick \citep[see e.g.][]{estrela_temperature_2022}. From iteration 3 to 4, the transit depth differs by roughly 30~ppm for wavelengths between 0.7~$\mu$m to 10~$\mu$m. For the wavelength range of MIRI LRS, the differences between iterations is always below the spectral accuracy even if 10 transits are observed. From iteration 4 to 5, the cloud top again changes but by less than 25~ppm. The residuals between iteration 4 to 5 are now consistently below the spectral precision of NIRSpec Prism, as illustrated by the black data points. The only exception to this are the CH$_4$ features around 2~$\mu$m to 4~$\mu$m, that still have residuals around 30~ppm. Since the differences between the transmission spectra of iteration 3, 4, and 5 are close to or below the spectral precision of JWST NIRSpec Prism, we stop our iterative procedure after iteration 5.
    
    Furthermore, comparing the residuals to the differences between the morning and evening terminator (see Fig.~\ref{fig:gcm_spectra}) shows that the differences between the limbs are generally larger then the residuals between iterations 4 an 5. This holds true for the cloud continuum (limb $\approx$~250~ppm versus residuals < 30 ppm), the methane features (limbs $\approx$~200~ppm versus residuals $<$~40~ppm), and the water features (limb $\approx$~100~ppm versus residuals $\approx$~30~ppm).

    \section{The combined 3D model for the warm Saturn HATS-6b that orbits an M-dwarf}
    \label{sec:sim}

    \begin{figure*}
        \centering
        \includegraphics[width=0.9\hsize]{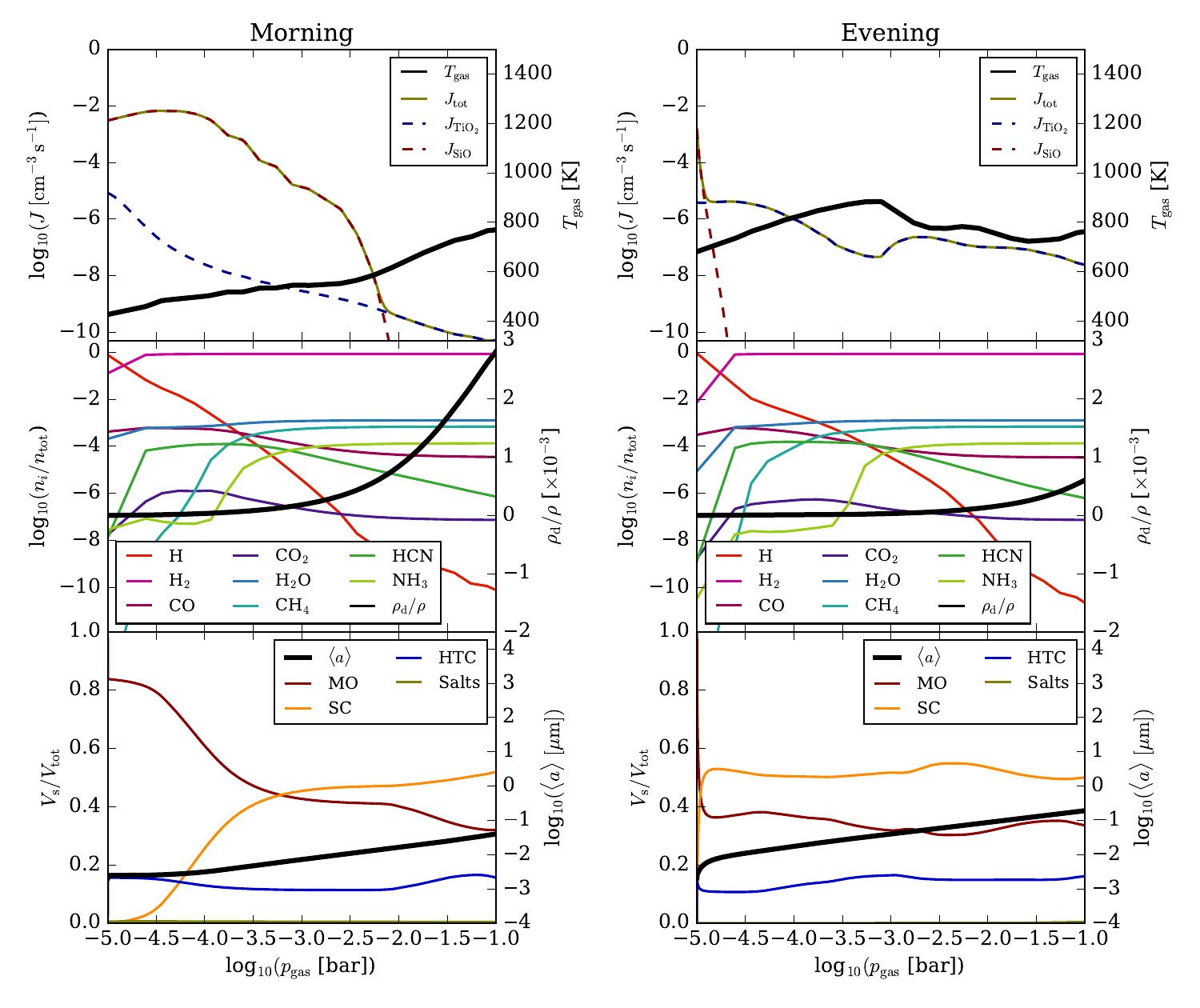}
        \caption{Cloud structure of the morning (\textbf{left}) and evening (\textbf{right}) terminator of HATS-6b. \textbf{Top:} gas temperature $T_\mathrm{gas}$ and total $J_\star$ and $J_\mathrm{TiO_2}$, and $J_\mathrm{SiO}$. \textbf{Middle:} gas-phase relative concentrations $n_i/n_\mathrm{gas}$ and cloud mass fraction $\rho_\mathrm{d} / \rho$. \textbf{Bottom:} cloud particle volume fractions $V_\mathrm{s} / V_\mathrm{tot}$ and mean cloud particle size $\langle a \rangle$  (MO = Metal oxides, SC = Silicates, HTC = High temperature condensates). Salts are a minor component and have a volume fraction close to 0.}
        \label{fig:sim_hats6b_overview}
    \end{figure*}
    
    \begin{figure*}
        \centering
        \includegraphics[trim={0cm 0cm 0cm -1.5cm}, width=\hsize]{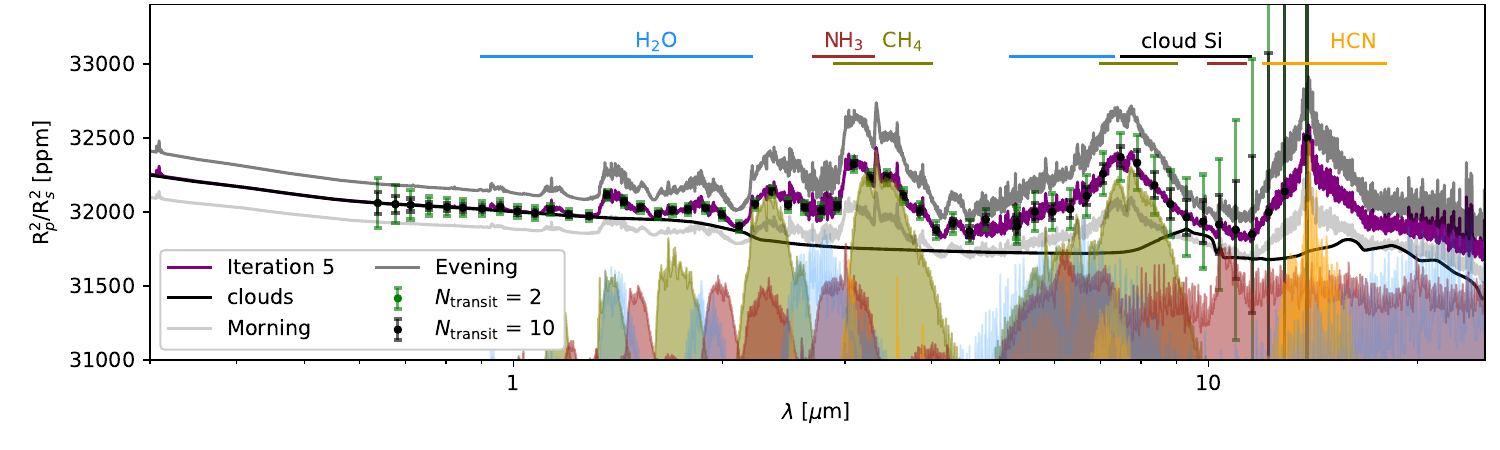}
        \caption{Transmission spectra for the warm Saturn HATS-6b for $\lambda$ = 0.3 $\mu$m to 25 $\mu$m including the total spectrum of the final iteration (iteration 5) as well as the morning and evening terminator separately. Also shown are the contributions of selected gas-phase specie, the contribution of the Si-bearing cloud particle materials, and the spectral accuracy of the JWST instruments NIRSpec Prism and MIRI LRS for 2 and 10 transits.}
        \label{fig:gcm_spectra}
    \end{figure*}

    The final combined 3D model for the warm Saturn HATS-6b that orbits an M-dwarf has been demonstrated in Sect~\ref{sec:res}. The two isobar maps at pressure ranges that are accessible to transmission spectroscopy are shown in the bottom right panel of Fig.~\ref{fig:gcm_slices} (iteration 5). HATS-6b is affected by a strong equatorial jet that reaches down to depth of $p_{\rm gas}\approx 1$bar. Horizontally converging flows\footnote{ Convergence is here defined as $\mathbf{\nabla} \cdot \mathbf{v} <$ 0, where $\mathbf{v}$ is the horizontal velocity vector. } further contribute to the global mixing of the atmosphere in the pole regions \citep[e.g.][]{baeyens_grid_2021}. The temperature inversion is well represented in the upper atmosphere with a temperature difference of $\approx$400K as compared to lower regions where $p_{\rm gas}\approx 10^{-2}$ bar. Sect.~\ref{sec:dis_hats6b} presents the final results from the combined model and Sect.~\ref{sec:sim_spectra} shows the transmission spectrum as observable with JWST NIRspec Prism and MIRI LRS.

    \subsection{The atmospheric structure of HATS-6b}
    \label{sec:dis_hats6b}

    HATS-6b is a warm Saturn with a hydrogen dominated atmosphere that is assumed to be oxygen rich in this study. Figure~\ref{fig:sim_hats6b_overview} presents the solution of the combined 3D model. The morning (left) and the evening (right) 1D terminator profiles were selected to demonstrate the cloud and kinetic chemistry results in enough detail to gain insights into the physical reasons for the differences between the respective transmission spectra that are shown in Sect.~\ref{sec:sim_spectra}.
    
    Figure~\ref{fig:sim_hats6b_overview} (top) shows that the local gas temperatures differ between the morning and the evening terminator which has direct consequences for the cloud formation efficiency. By comparing the nucleation rates ($J_*$ [cm$^{-3}$s$^{-1}$]), we notice that the cooler morning terminator (left) forms more cloud particles than the hotter evening terminator. Consequently, the mean particles size, $\langle a \rangle$ [$\mu$m], is smaller in the respective low gas pressure region of the morning terminator. The lower panels in Fig.~\ref{fig:sim_hats6b_overview} show which material groups dominate the cloud particle composition, and hence, their opacities. Fig.~\ref{fig:app_matcomp} contains the detailed breakdown of all 16 modelled bulk materials. Materials that are determined by low-abundance elements like Cl, Mn etc. can be thermally stable but appear with $\sim$1\% volume contribution \citep{helling_dust_2017, woitke_equilibrium_2018}.

    Figure~\ref{fig:sim_hats6b_overview} demonstrates that it is reasonable to expect the cloud particles that form in the atmosphere of a warm Saturn like HATS-6b to be made of a mix of materials and that their mean particles sizes changes throughout the atmosphere. It further demonstrates that the bulk cloud mass ($\rho_\mathrm{d} / \rho$, black line, middle panel) is not necessarily equally distributed through the atmospheres  and differs between the two terminators, hence, emphasising the terminator asymmetry. Comparing to the material volume fractions, $V_\mathrm{s} / V_\mathrm{gas}$, in the lower panels demonstrates that the bulk mass comes from silicates (MgSiO$_3$[s], Mg$_2$SiO$_4$[s], CaSiO$_3$[s], Fe$_2$SiO$_4$[s]) and metal oxides (SiO[s], SiO$_2$[s], MgO[s], FeO[s], Fe$_2$O$_3$[s]) because Mg/Si/Fe/O are the most abundant elements amongst those considered here. It is, hence, the result of mass conservation. Far less important for the cloud material composition are the high temperature condensates (TiO$_2$[s], Al$_2$O$_3$[s]) and salts (KCl[s], NaCl[s]) for the same reason. Since most dominant condensation species are oxygen bearing species and oxygen is far more abundant then Ti, Mg, Si, and Fe, these condensation species do not significantly impact the gas-phase chemistry of non Ti, Mg, Si, and Fe bearing species.

    As introduced in Sect. \ref{sec:sim_chem}, the disequilibrium chemistry of HATS-6b is shown in Figure~\ref{fig:abun}. The middle and bottom panels show the relative concentrations for a selection of molecular gas species for the last iteration at four coordinates along the equator of the planet: sub-stellar point, anti-stellar point, and the two terminators. The variations along the equator are caused both by differences in the ($T_{\rm gas}$, $p_{\rm gas})$ profiles for the four coordinates, and by differences in the stellar XUV radiation and the SEPs. The sub-stellar point is irradiated both by XUV radiation and SEPs, the terminators are irradiated only by SEPs, and the anti-stellar point is irradiated by neither. Comparing the four cases it can be seen that the sub-stellar point and the two terminators are relatively similar, whereas the anti-stellar point differs significantly from the rest. By comparing the sub- and anti-stellar point, we notice a steep decrease in the concentration of many of the gas species at the sub-stellar point (incl. H$_2$, CH$_4$, and NH$_3$) in the upper atmosphere, indicating a break down of these molecules through photolysis by the XUV radiation or ionisation by the SEPs. Other species (such as H, HCN, and partly CO and CO$_2$) show an increase in concentration for the sub-stellar point compared to the anti-stellar point as we move further up into the atmosphere, indicating that these are positively influenced by photochemical reactions. Comparing model runs for the sub-stellar point with and without SEPs (middle panel), we notice a significant contribution by the SEPs on the gas-phase concentrations and that this effect reaches far down into the atmosphere. The middle and bottom panels show that the sub-stellar point with SEPs bear a strong resemblance to the terminator regions, which could indicate that SEPs can have a larger effect than the XUV radiation on the concentrations of the gas species, including observationally interesting species such as CH$_4$ and HCN. The bottom panel shows that the differences between the terminator regions are generally higher at lower pressures, with the exception of H that also show differences deeper into the atmosphere.

    As mentioned in Sect. \ref{sec:Introduction}, M-dwarf stars such as HATS-6 are known to have higher magnetic activities compared to more massive stars like the Sun. This increased activity leads to increased amounts of SEPs. As explained in Sect. \ref{sec:Model_argo}, the amount of SEPs has been found to scale with the X-ray flare intensity of the star, which for M-dwarf stars has been reported to range from 0.001 to 0.2 Wm$^{-2}$ at 1AU. In this study we scale our SEP spectrum based on a X-ray flare intensity of 0.1 Wm$^{-2}$ at 1AU, indicating that the amount of SEPs could be significantly higher than what we show, leading to a greater impact on the disequilibrium gas-phase chemistry.

    \subsection{Transmission spectra}
    \label{sec:sim_spectra}

    Transmission spectra are used to study the chemical composition and atmospheric asymmetries of gaseous planets. To determine the concentrations of known non-equilibrium species (e.g. CH$_4$ and HCN), the H/C/N/O complex of the gas-phase is determined using ARGO (Sect.~\ref{sec:sim_chem}). As the methods used in this study may help to efficiently interpret the data of JWST and future space missions, the focus was set on spectral features relevant for the JWST instruments NIRspec and MIRI. In the following, the near-IR transmission spectra that also include non-equilibrium species like CH$_4$ and HCN for a wavelength range of $\lambda$ = 0.3 $\mu$m to 12 $\mu$m are presented.

    To create the transmission spectrum of HATS-6b, we used the ($T_{\rm gas}$, $p_{\rm gas})$ profiles from the GCM, the cloud opacities from the kinetic cloud model and the disequilibrium chemistry from ARGO. The full transmission spectra is shown in Fig. \ref{fig:gcm_spectra}. Also shown are the morning and evening spectra separately as well as the cloud and the gas-phase contributions. The transmission spectrum shows a roughly flat spectra for wavelengths below 2~$\mu$m, owing to scattering which is typical for cloudy exoplanets \citep{bean_ground-based_2010, crossfield_warm_2013, kreidberg_clouds_2014, knutson_hubble_2014, sing_hst_2015, sing_continuum_2016, benneke_sub-neptune_2019}. Observing HATS-6b at these wavelengths would allow to determine the height of the cloud deck \citep{mukherjee_cloud_2021}. Above 3~$\mu$m the gas-phase species start to contribute significantly to the spectrum. Especially CH$_4$ (around 3~$\mu$m and 8~$\mu$m), H$_2$O (around 6~$\mu$m), and HCN (around 10.5~$\mu$m) can be seen. However, the clouds still heavily mute the spectral features, making them harder to detect.

    Around 10~$\mu$m, the cloud opacities show a broad silicate feature from the Si-bearing cloud particle materials. This feature originates from Si-O vibrations \citep{wakeford_transmission_2015}. Detecting this feature could confirm the presence of silicon-bearing cloud particle materials in HATS-6b as has been done previously for WASP-17b \citep{grant_jwst-tst_2023} and WASP-107b \citep{dyrek_so2_2023}.

    HATS-6b itself may reveal differences in cloud coverage and chemistry between the morning and evening terminator similar to WASP~39b \citep{carone_wasp-39b_2023, espinoza_inhomogeneous_2024}. The cloud deck is slightly higher at the evening terminator compared to the morning terminator leading to a $\approx$~250~ppm offset at the short wavelengths (0.3~$\mu$m < $\lambda$ < 2~$\mu$m). The observable difference in the gas-phase is mainly caused by CH$_4$ which has $\approx$~200 ppm higher signal at the evening terminator than at the morning terminator. H$_2$O has $\approx$~100~ppm higher signal in the evening terminator than the morning terminator. At the morning terminator, the contribution of chemical species to the transmission spectrum can vary strongly with latitude (see Fig.~\ref{fig:app_allspec}). Especially at $\pm 45^\circ$ of the morning terminator, stronger transmission features of CH$_4$ and HCN arise as a consequence of the Rossby gyres that represent particularly cold regions in the atmosphere (see Fig.~\ref{fig:gcm_slices}).  Warm Saturns and Jupiters in the temperature range between 800~K and 1200~K typically exhibit extended Rossby wave gyres at the morning terminator \citep[e.g.][Fig. 2]{baeyens_grid_2021}. Thus, a study of morning terminator chemistry in warm Saturns needs to take into account special locations like the dynamically cool Rossby gyres to correctly infer global atmosphere properties.

\section{Discussion}
\label{sec:Discussion}
    
    In this work, a 3D climate  model in combination with a microphysical cloud model was implemented to capture the feedback between the heating, the 3D climate, and cloud formation in its full complexity. The model was applied to the particular example of a warm Saturn around an M-dwarf, HATS-6b. Many exoplanet theories predict temperature inversions in exoplanet atmospheres due to various reasons. In Sect.~\ref{sec:dis_antig}, the strong cloud-induced temperature inversion of HATS-6b is discussed. Furthermore, we discuss the dynamics of the terminators in Sect.~\ref{sec:dis_dyninst}. While several studies have modelled the climate and cloud structure of warm Saturns, there have been no detailed studies of warm Saturns around M-dwarf stars like HATS-6b so far. Therefore in Sect.~\ref{sec:dis_comp}, the results for HATS-6b are compared to grid studies including warm Saturns and detailed models of other Saturn-mass planets similar to HATS-6b.

    \subsection{Anti-greenhouse effect}
    \label{sec:dis_antig}
    
    The clouds in HATS-6b have considerable cloud particle sizes and number densities for pressures lower than 10$^{-3}$ bar. The high cloud deck leads to scattering and absorption of the incoming short wavelength radiation on cloud particles. This reduces the radiative heating of the lower atmospheric layers while simultaneously heating up the upper layers. Both effects lead to a temperature inversion in the layers where clouds are located and a cooling of the layers below. This effect is called the anti-greenhouse effect and was first observed in Titan \citep{mckay_greenhouse_1991}. It has also been theoretically predicted for exoplanet atmospheres with either an extended cloud deck or photochemical hazes \citep{heng_effects_2012, morley_neglected_2012, steinrueck_photochemical_2023}.

    In hotter exoplanets, temperature inversions are not caused by scattering from high-altitude clouds but by gas-phase species like TiO and VO \citep{hubeny_possible_2003, fortney_unified_2008} or AlO, CaO, NaH, and MgH \citep{gandhi_new_2019}. These species heat up the upper atmosphere due to very efficient absorption of stellar radiation in the optical wavelength range. These absorption-driven temperature inversions have been observed in (ultra) hot Jupiters \citep{haynes_spectroscopic_2015, yan_temperature_2020, yan_detection_2022}. HATS-6b orbits an M-dwarf that emits less flux in the optical wavelength range than a hotter star \citep{lothringer_influence_2019}. Furthermore, \citet{hu_radiative_2011} predicted that an exoplanet with a \ce{CO2} dominated atmosphere orbiting an M-dwarf stars would also have an anti-greenhouse effect mainly due to highly efficient Rayleigh scattering of \ce{CO2} at the top of the atmosphere. In this study, however, \ce{CO2} is a minor species and Rayleigh scattering is dominated by hydrogen and helium. In our simulation of HATS-6b, no temperature inversion is present in the GCM runs without clouds. Therefore, we are confident in our conclusion that the temperature inversion is indeed caused mainly by scattering of stellar irradiation at the top of the extended cloud deck similar to the anti-greenhouse effect observed in Titan due to scattering at the top of the atmosphere due to hazes.
    
    To observe a temperature inversion directly, emission from the lower atmosphere needs to be observable \citep{gandhi_new_2019}. The strong cloud coverage will likely block all emissions from the lower atmosphere and make direct detections of the temperature inversion unlikely. However, HATS-6b lies in a temperature regime where quenching is expected to affect the chemistry \citep[e.g.][]{baeyens_grid_2021}. Fig.~\ref{fig:gcm_residuals} and \ref{fig:gcm_spectra} shows a CH$_4$ feature that is visible despite the cloud layer, which indicates that the CH$_4$ extends into the upper layers of the atmosphere where it is exposed to photolysis and high degrees of SEPs. Since CH$_4$ is very susceptible to photo-chemical reactions \citep{moses_disequilibrium_2011, line_thermochemical_2011, baeyens_grid_2021, baeyens_grid_2022, konings_impact_2022}, it will most likely break down in the upper atmosphere, and in order to maintain a visible CH$_4$ feature, a constant up-welling from the lower protected part of the atmosphere would be necessary. A visible CH$_4$ feature could therefore indicate that vertical mixing has connected the observable gas-phase chemistry above the clouds to deeper atmosphere layers, thereby forming a probe through the temperature inversion and into the layers cooled by the anti-greenhouse effect \citep{agundez_pseudo_2014, fortney_beyond_2020}. In addition, this work confirmed the influence of stellar energetic particles (SEP) on the day side chemistry of a planet around a relatively active planet as was already pointed out by \citet{barth_moves_2021} for HD~189733b.

    \subsection{Dynamics of the terminator regions}
    \label{sec:dis_dyninst}

\begin{figure}
        \centering
        \includegraphics[trim={0cm 0cm 0cm 0cm}, clip, width=0.95\hsize]{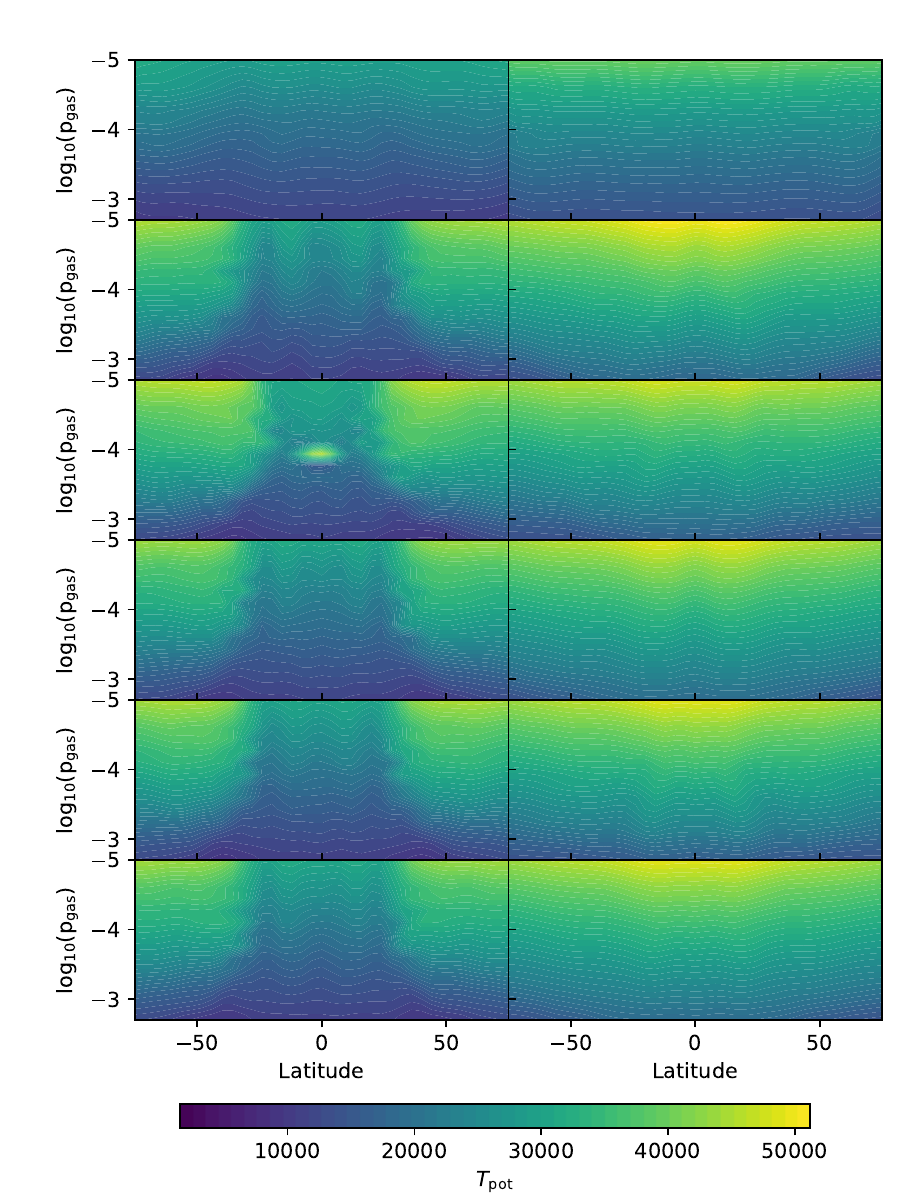}
        \caption{Potential temperature cross section across the morning (left panels) and evening terminators (right panels) for iterations 0 to 5 (from top to bottom). }
        \label{fig:instability}
    \end{figure}

    In Sect.~\ref{sec:res}, we seen that the morning terminator of iteration~2 develops a temperature spike in the upper atmosphere (see Fig.~\ref{fig:gcm_tp}). Since no other iteration shows such a spike, we can conclude that it is caused by the impact of the specific configuration of the static cloud opacities of iteration~2 on the temperature and wind structure of HATS-6b. A closer inspection of the potential temperature profiles across the morning and evening terminators of all iterations (Fig.~\ref{fig:instability}) reveals a potential temperature anomaly at the equator for the morning terminator in iteration~2. The potential temperature is linked to atmospheric stability. If the potential temperature increases monotonically with height, the atmosphere is dynamically stable \citep[e.g.][]{holton_introduction_2013}.
    
    The equatorial morning terminator is a region, where the super-rotating wind jet advects colder air from the night side to the day-side that is heated by the cloud feedback (see Fig. \ref{fig:gcm_slices}). The advection of cold air results in a minimum in potential temperature between -40 degree and +40 degree latitude. In contrast to that, the potential temperature across the evening terminator is more barotropic. Here, the wind jet broadens and encompasses almost the whole evening terminator region (Fig. \ref{fig:gcm_slices}), resulting in a more uniform advection of warm air towards the night side.
    
    The aim of this work is to identify a stable climate solution for a cloudy warm Saturn orbiting an M-dwarf star. However, it may be worthwhile investigating if the strong temperature gradients of the morning terminator at the boundary of the super-rotating jet give rise to instabilities and thus variations at the morning terminators as is evident in iteration~2. Observations that aim to disentangle the atmospheric properties of the morning and evening terminator of gas-giants with JWST could confirm the stronger dynamical variability of the morning terminator compared to the evening terminator. However for HATS-6b, we do not see observable differences due to variations across the terminators.

    \subsection{Comparison to other models}
    \label{sec:dis_comp}

    The cloud structure and climate of HATS-6b was simulated as a first example of a warm Saturn around an M-dwarf host star. To extend our findings from HATS-6b to general warm Saturns, the results are compared to other studies focusing on including the 3D cloud and climate of warm Saturn type exoplanets. 
    
    \citet{helling_exoplanet_2023} conducted a grid study of post-processed cloud structures for temperatures between 400~K and 2600~K for F, G, K and M-dwarf stars. For exoplanets with an equilibrium temperature of 700~K around M-dwarf stars, they predict a strong uniform cloud coverage. The equilibrium temperature and cloud structure of HATS-6b falls with their "class (i)" planets which are characterised by global and mostly homogenous cloud coverage. Our results suggest that at least for exoplanets around M-dwarf stars like HATS-6b an anti-greenhouse effect is expected for these class (i) planets caused by the clouds.
    
    \citet{christie_impact_2022} studied the impact of clouds on the climate of the warm Neptune GJ~1214b. They considered phase equilibrium clouds within the \texttt{EddySed} model \citep{ackerman_precipitating_2001}. Their cloud model parameterises the settling of cloud particles with the parameter $f_\mathrm{sed}$, where a lower $f_\mathrm{sed}$ results in vertically extended clouds. In contrast to our work, and matching other studies of cloud composition of GJ~1214b \citep{gao_microphysics_2018, ormel_arcis_2019}, they consider only KCl and ZnS as cloud particle material. \citet{christie_impact_2022} found that clouds can cause cooling in the lower atmosphere ($10^{-2}$ bar < $p$ < 1 bar) which matches our findings. This temperature decrease is most pronounced for higher metallicities and lower $f_\mathrm{sed}$. However, they did not find a significant temperature increase in the upper atmosphere, as we find in our work.

    Another planet similar to HATS-6b, is the Saturn-mass exoplanet WASP-39b that is part of the JWST Early Release Science program \citep{feinstein_early_2023, ahrer_early_2023, rustamkulov_early_2023, alderson_early_2023, jwst_transiting_exoplanet_community_early_release_science-team_identification_2023}. WASP-39b is at the upper limit of the class(i) cloud temperature regime identified by \citet{helling_exoplanet_2023} with $T_{\rm eq} \sim 1100~{\rm K}$. Cloudless GCMs of WASP-39b show relatively small day-night temperature differences of $\Delta T\sim500~{\rm K}$ \citep{carone_wasp-39b_2023,lee_mini-chemical_2023}, similar to the cloudless simulations of HATS-6b. Post-processed cloud modelling by \citet{carone_wasp-39b_2023} predicts global cloud coverage of WASP-39b. Pre-JWST observations pointed towards a relatively cloud free atmosphere, with estimates of atmospheric metallicities ranging from $0.1\times - 100\times$ solar \citep{sing_continuum_2016, nikolov_vlt_2016, fischer_hst_2016, wakeford_complete_2018}. Observations with JWST revised these earlier observational results and indicate the presence of clouds and a $\gtrsim10\times$ solar metallicity, with some models favouring inhomogeneous cloud coverage \citep{feinstein_early_2023}. Thus, JWST observations of WASP-39b demonstrated that it is possible to break the high metallicity and cloudiness degeneracy that also plagued warm Saturns pre-JWST \citep{carone_indications_2021}. JWST observations of WASP-39b could reveal cloud asymmetries between the morning and evening terminator as predicted by \citet{carone_wasp-39b_2023}. Also in this work, we find a tendency for cloud asymmetries between both terminators (see Sect.~\ref{sec:sim_spectra}.)

\section{Conclusion}
\label{sec:conclusion}

    In this paper, we explored the atmospheric, micro-physical cloud and gas-phase structure of the warm Saturn HATS-6b, orbiting an M-dwarf using a combined model in the form of step-wise iterations between a detailed cloud formation description and \texttt{expeRT/MITgcm}, a 3D GCM with full radiative transfer and deep atmosphere extension. We demonstrated how the combined modelling approach can help to support the interpretation of the data from space missions (e.g. JWST) for warm Saturn type planets.

    We find a significant cloud coverage on both the day and night side of HATS-6b, which is to be expected for the generally low temperatures (500 K < T < 1200 K) of warm Saturn type exoplanets. On the day-side and evening terminator, we find that clouds cause a significant temperature inversion in the upper atmosphere ($p$ < $10^{-2}$ bar). This also results in a globally reduced temperature of the deeper atmospheric layers ($10^{-2}$ bar < $p$ < 1 bar) which is consistent with an anti-greenhouse effect. Including clouds in the GCM also leads to a stronger and narrower equatorial jet stream. The transmission spectra of HATS-6b shows a characteristically flat spectrum in the optical. In the infrared, however, major molecular absorption features like \ce{CO2, CH4} and \ce{H2O} are visible despite the global cloud coverage. Around 10 $\mu$m, a silicate cloud feature is predicted. The small radii of M-dwarf stars compared to more massive stars leads the to a 'magnifying effect' of spectral features within transmission spectra. This makes planets like HATS-6b prime targets for deciphering gas chemistry and cloud compositions for warm Saturn type exoplanets. For wavelengths up to 8 $\mu$m, it may even be possible to identify morning and evening terminator differences with the CH$_4$ feature around 3 $\mu$m.

    When iterating between the GCM and the cloud model, the differences in the atmospheric structure of HATS-6b quickly drop below the observational accuracy of the JWST NIRSpec Prism and MIRI LRS. For HATS-6b this was reached after 5 iterations. Even after 1 iteration (GCM $\rightarrow$ Clouds $\rightarrow$ GCM) the GCM predicts an atmospheric structure that differs by less than 50ppm to iteration 5. Furthermore, differences in the transmission spectra between the iteration steps are smaller than the differences between the morning and evening terminator. Therefore a combined model that enables the full complexity of all modelling components (here 3D GCM and cloud formation) proves useful for the scientific interpretation of observational data for larger sets of exoplanets to be studied with CHEOPS, JWST, and also ELT, PLATO, and Ariel in the future.

    Warm Saturn type exoplanets around M-dwarfs are ideal candidates to identify cloud particle composition by observing their spectral features and identify the cloud-induced thermal inversion that arises on these planets. In particular limb asymmetry studies will allow us to study differences in the chemistry and cloud top which are caused by the feedback of clouds on the atmospheric structure. Using the combined model allows us to account for the full physical complexity of each model in a computationally feasible manner and enables a detailed interpretation of observational data within the accuracy of JWST NIRSpec and MIRI LRS.

\begin{acknowledgements}
    S.K. N.BM., A.D.S, F.A, H.LM., L.C., Ch.H., L.D., and U.G.J. acknowledge funding from the European Union H2020-MSCA-ITN-2019 under grant agreement no. 860470 (CHAMELEON). 
    D.S. and D.A.L. acknowledge financial support and use of the computational facilities of the Space Research Institute of the Austrian Academy of Sciences. We acknowledged the computation support at CPH and at the IWF Graz through the Vienna Science Cluster (VSC project 72245). LD acknowledges support from the KU Leuven IDN grant IDN/19/028 grant Escher. U.G.J. further acknowledges funding from the Novo Nordisk Foundation Interdisciplinary Synergy Programme grant no. NNF19OC0057374. To achieve the scientific results presented in this article we made use of the Python, especially the NumPy \citep{harris_array_2020} and Matplotlib \citep{hunter_matplotlib_2007}. The post-processing of GCM datawas performed with gcm-toolkit \citep{schneider_gcm_toolkit_2022}.
\end{acknowledgements}

\bibliographystyle{aa}
\bibliography{bibliography}

\begin{appendix}

   \section{Additional cloud structure plots}
   \label{sec:App_matcomp}

\begin{figure*}
        \centering
        \includegraphics[trim={0cm 0cm 0cm 0cm}, clip, width=0.92\hsize]{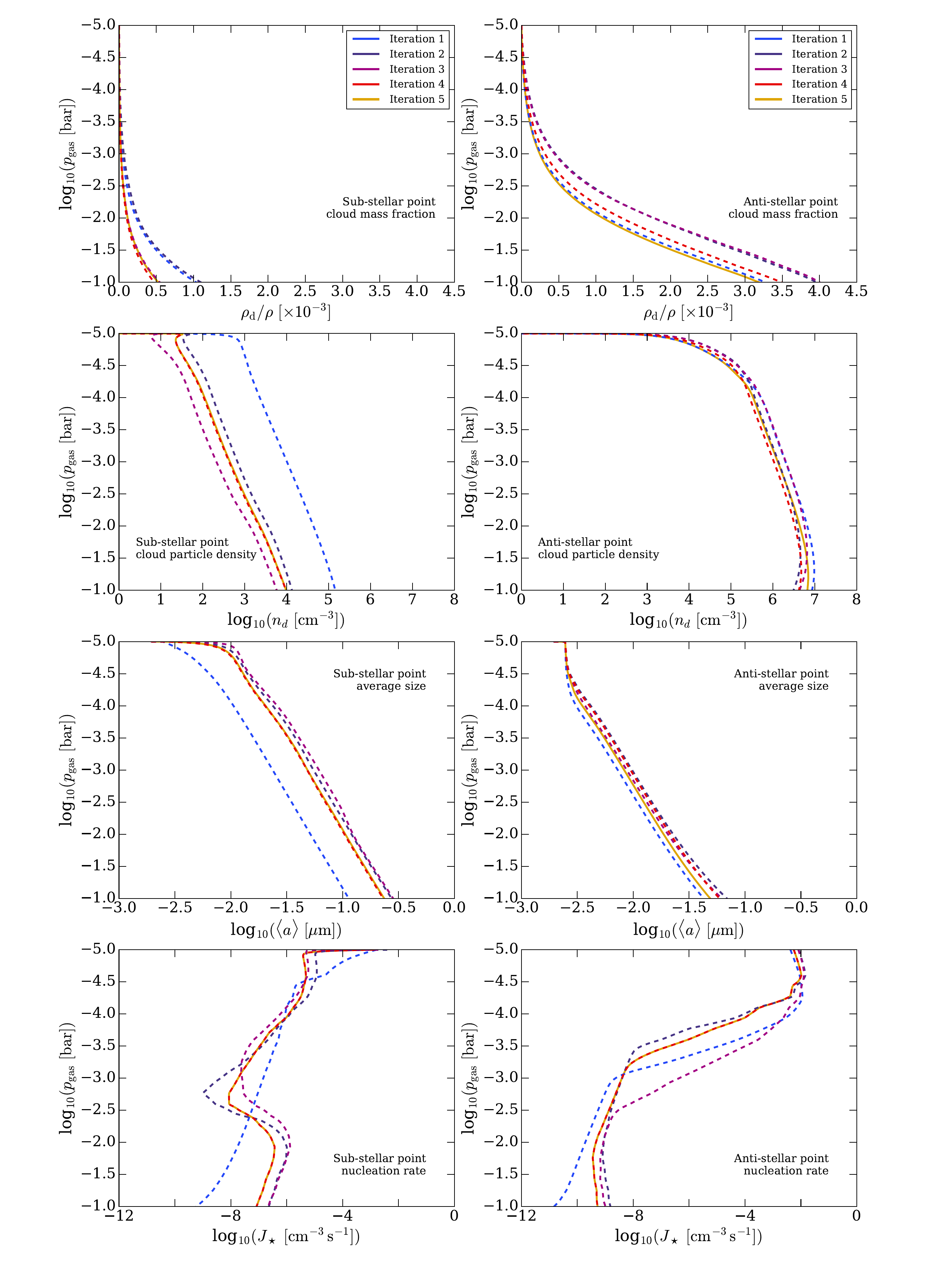}
        \caption{Cloud particle properties of the step-wise iterated cloud structure for the warm Saturn example HATS-6b. Iteration 5 is shown as a solid line to highlight the final result. \textbf{Left:} sub-stellar point. \textbf{Right:} anti-stellar point. \textbf{Top:} cloud mass fraction $\rho_\mathrm{d}/\rho$. \textbf{Upper middle:} cloud particle number density $n_d$. \textbf{Lower middle:} average cloud particle size $\langle a \rangle$. \textbf{Bottom:} Nucleation rate $J_\star$. The morning and evening terminator can be found in Fig~\ref{fig:cloud_iterations}.}
        \label{fig:cloud_iterations_stellar}
    \end{figure*}

\begin{figure*}
        \centering
        \includegraphics[trim={0cm 0cm 0cm 0cm}, clip, width=\hsize]{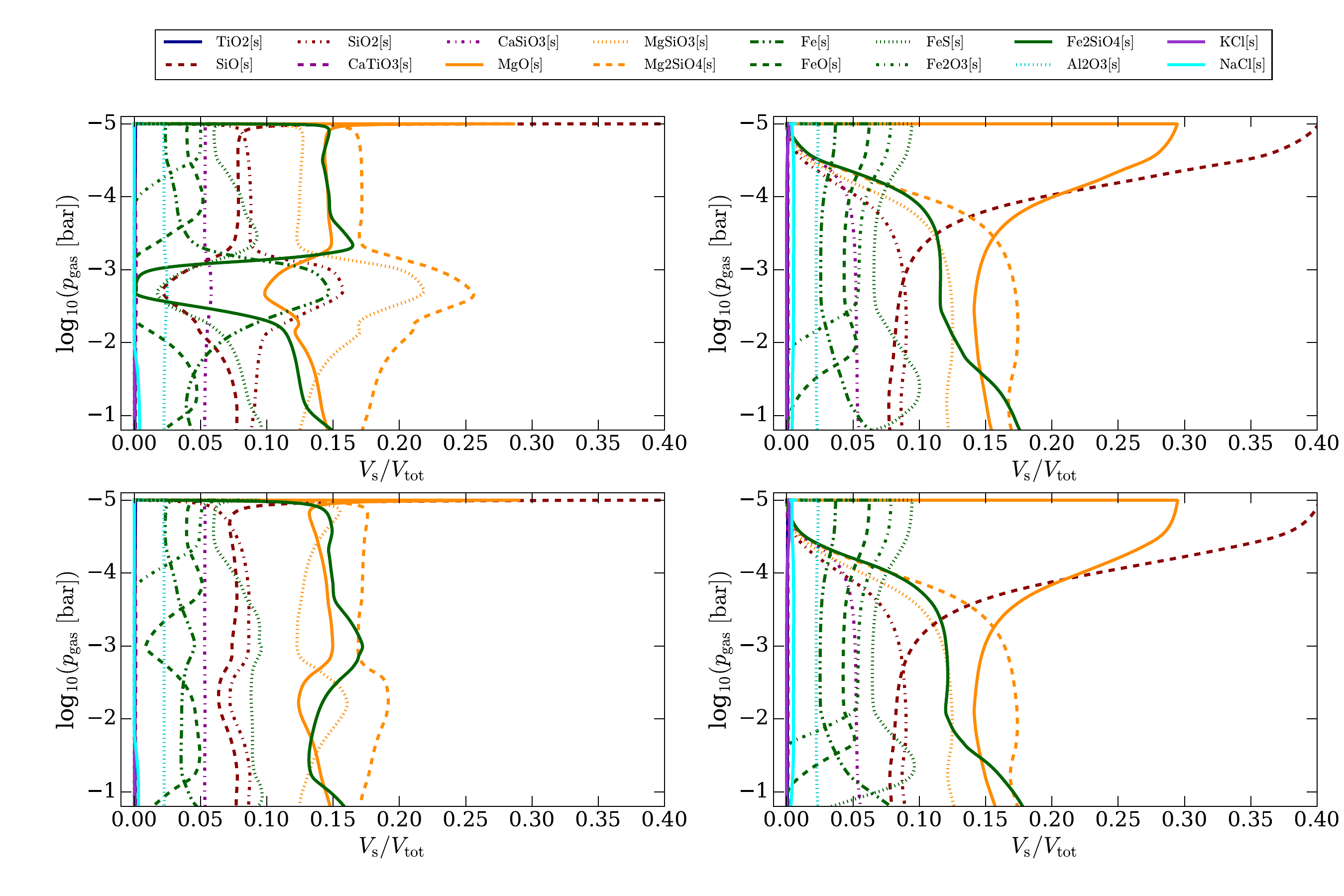}
        \caption{Cloud particle material composition of iteration 5. \textbf{1st row:} Sub-stellar point. \textbf{2nd row:} Anti-stellar point. \textbf{3rd row:} Evening terminator. \textbf{4th row:} Morning terminator. Here the cloud profiles only reach down to 0.3~bar. }
        \label{fig:app_matcomp}
    \end{figure*}

\begin{figure*}
        \centering
        \includegraphics[width=\hsize]{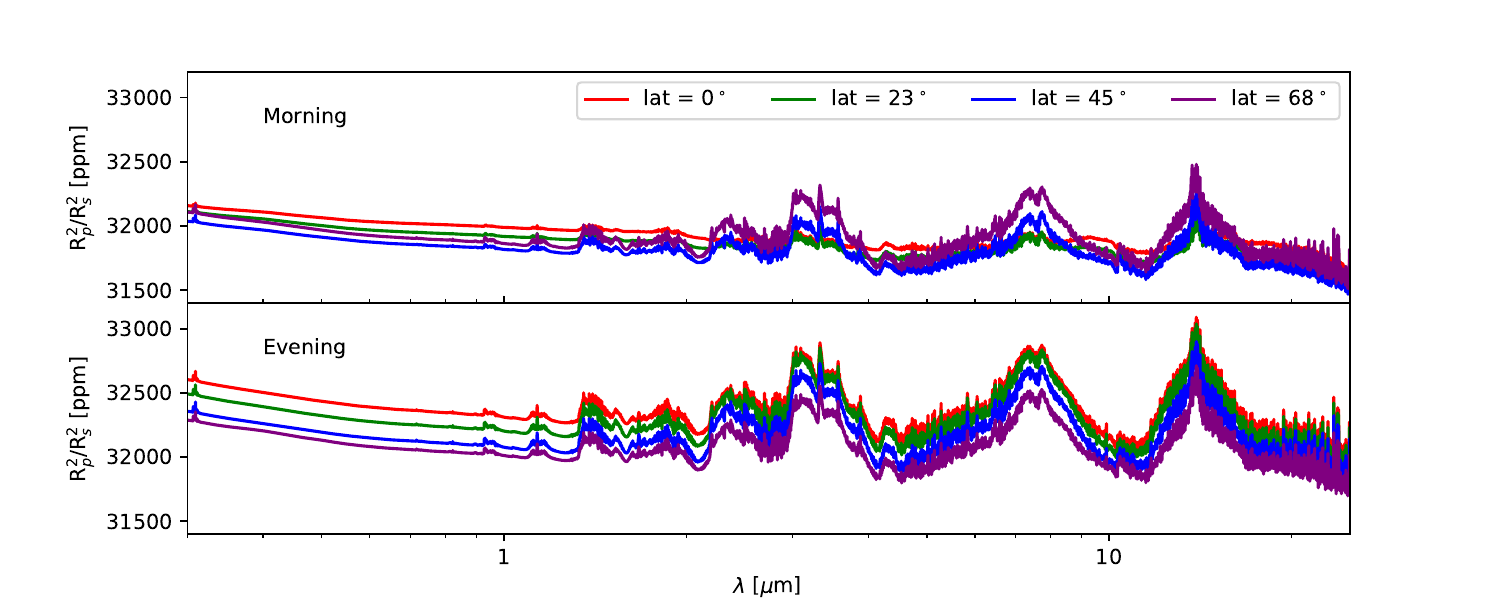}
        \caption{Transmission spectrum for each terminator grid cell. \textbf{Top:} Morning terminator. \textbf{Bottom:} Evening terminator.}
        \label{fig:app_allspec}
    \end{figure*}

    The cloud particle properties of the sub-stellar and anti-stellar point for all iterations are shown in Fig.~\ref{fig:cloud_iterations_stellar}. Additionally for iteration 5, the cloud particle material composition of the sub-stellar point, anti-stellar point, morning terminator and evening terminator are shown in Fig. \ref{fig:app_matcomp}. Lastly, the transmission spectrum of each terminator grid cell (lat = \{ -68, -23, 0, 23, 68\}) of iteration 5 before averaging is shown in Fig.~\ref{fig:app_allspec}.

   \section{Opacity data}
   \label{sec:App_opadata}
   The references for the absorption cross sections used in the radiative transfer of the GCM and transmission spectra calculations are listed in Table~\ref{tab:Appendix_crosssecs}. The references for the cloud particle material opacities are listed in Table \ref{tab:Appendix_conductivities}. Due to missing data, CaSiO$_3$ is treated as vacuum.
   
    \begin{table}
        \centering
        \caption{References for cross sections of gas-phase species.}
        \label{tab:Appendix_crosssecs}
        \begin{tabular}{l l}
            \hline\hline
             Gas-phase species & Reference \\  
             \hline 
             H$_2$O & \citep{polyansky_exomol_2018} \\  
             CO$_2$ & \citep{yurchenko_exomol_2020} \\ 
             CH$_4$ & \citep{yurchenko_hybrid_2017} \\ 
             NH$_3$ & \citep{coles_exomol_2019} \\
             CO & \citep{li_rovibrational_2015} \\ 
             H$_2$S & \citep{azzam_exomol_2016} \\
             HCN & \citep{barber_exomol_2014} \\
             PH$_3$ & \citep{sousa-silva_exomol_2015} \\
             TiO & \citep{mckemmish_exomol_2019} \\
             VO & \citep{mckemmish_exomol_2016} \\
             FeH & \citep{wende_crires_2010} \\ 
             Na & \citep{piskunov_vald_1995} \\
             K & \citep{piskunov_vald_1995} \\
            \hline
        \end{tabular}
    \end{table}
    
    \begin{table}
        \centering
        \caption{References for the cloud material opacities.}
        \label{tab:Appendix_conductivities}
        \begin{tabular}{l l}
            \hline\hline
             Material species & Reference \\  
             \hline
            TiO$_2$[s] (Rutile) & \citet{zeidler_near-infrared_2011}  \\
            SiO$_2$[s] (alpha-Quartz) & \citet{palik_handbook_1985}, \citet{zeidler_optical_2013}  \\
            SiO[s] (polycrystalline) & Philipp in \citet{palik_handbook_1985} \\
            MgSiO$_3$[s] (grass) & \citet{dorschner_steps_1995} \\
            Mg$_2$SiO$_4$[s] (crystalline) & \citet{suto_low-temperature_2006} \\
            MgO[s] (Cubic) & \citet{palik_handbook_1985} \\
            Fe[s] (metallic) & \citet{palik_handbook_1985}  \\
            FeO[s] (amorphous) & \citet{henning_optical_1995} \\
            Fe$_2$O$_3$[s] (amorphous) & A.H.M.J. Triaud (unpublished) \\
            Fe$_2$SiO$_4$[s] (amorphous) & \citet{dorschner_steps_1995} \\
            FeS[s] (amorphous) & Henning (unpublished) \\
            CaTiO$_3$[s] (amorphous) & \citet{posch_infrared_2003}  \\
            Al$_2$O$_3$[s] (grass) & \citet{begemann_aluminum_1997}  \\
            KCl [s] (cubic) & \citet{palik_handbook_1985, querry_optical_1998} \\
            NaCl [s] (cubic) & \citet{palik_handbook_1985, querry_optical_1998} \\
            \hline
        \end{tabular}
    \end{table}

    \section{GCM convergence tests}
    \label{sec:App_gcm_convergence}

\begin{figure*}
        \centering
        \includegraphics[width=\hsize]{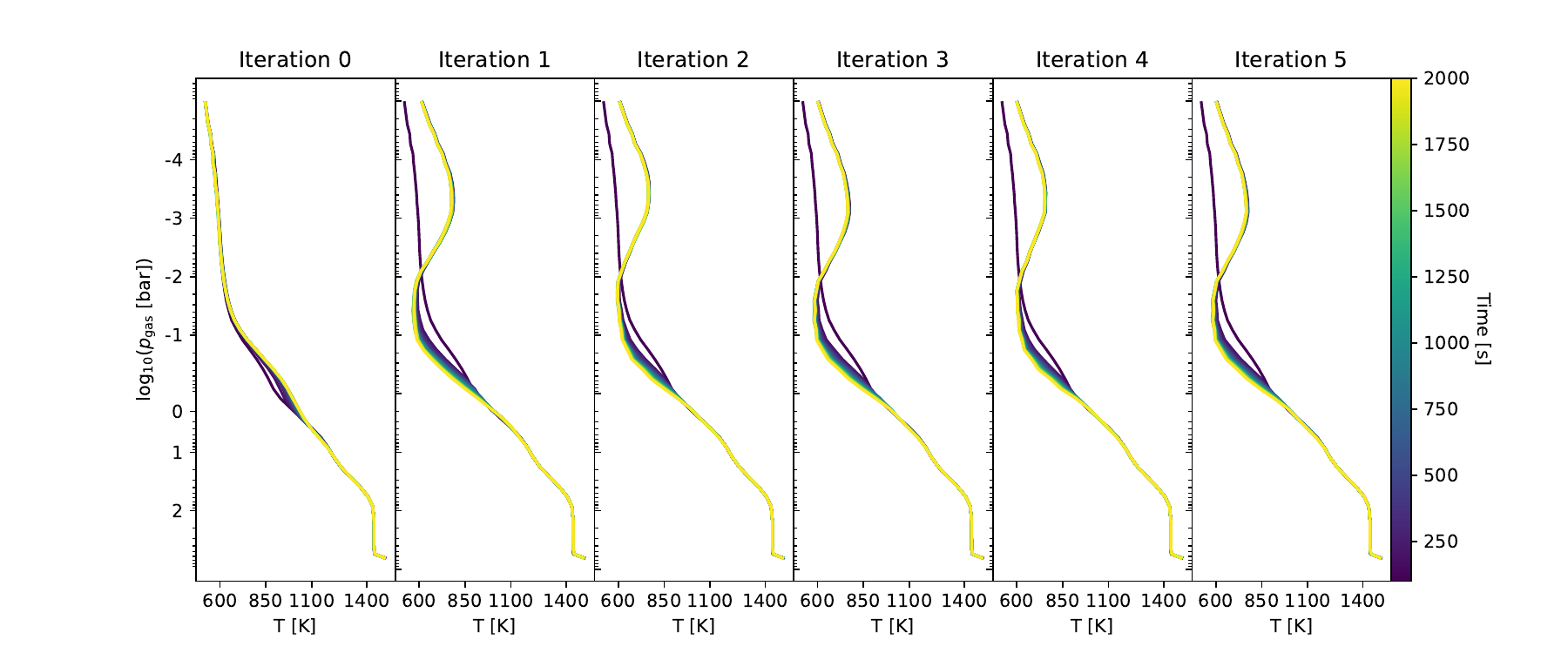}
        \caption{Evolution of the global average temperature for every 100 days of the \texttt{expeRT/MITgcm} runs.}
        \label{fig:app_gcm_teval}
    \end{figure*}

    To test the convergence of the \texttt{expeRT/MITgcm} runs, we analyse the time dependent temperature evolution of the GCM. The results can be seen in Fig. \ref{fig:app_gcm_teval}. The results show that the temperature structure of all three runs are reasonably converged in the upper layers of the GCM.

   \section{Input SED for ARGO}
   \label{sec:App_sed}
\begin{figure*}
        \centering
        \includegraphics[trim={0cm 0cm 0cm 0cm}, clip, width=\hsize]{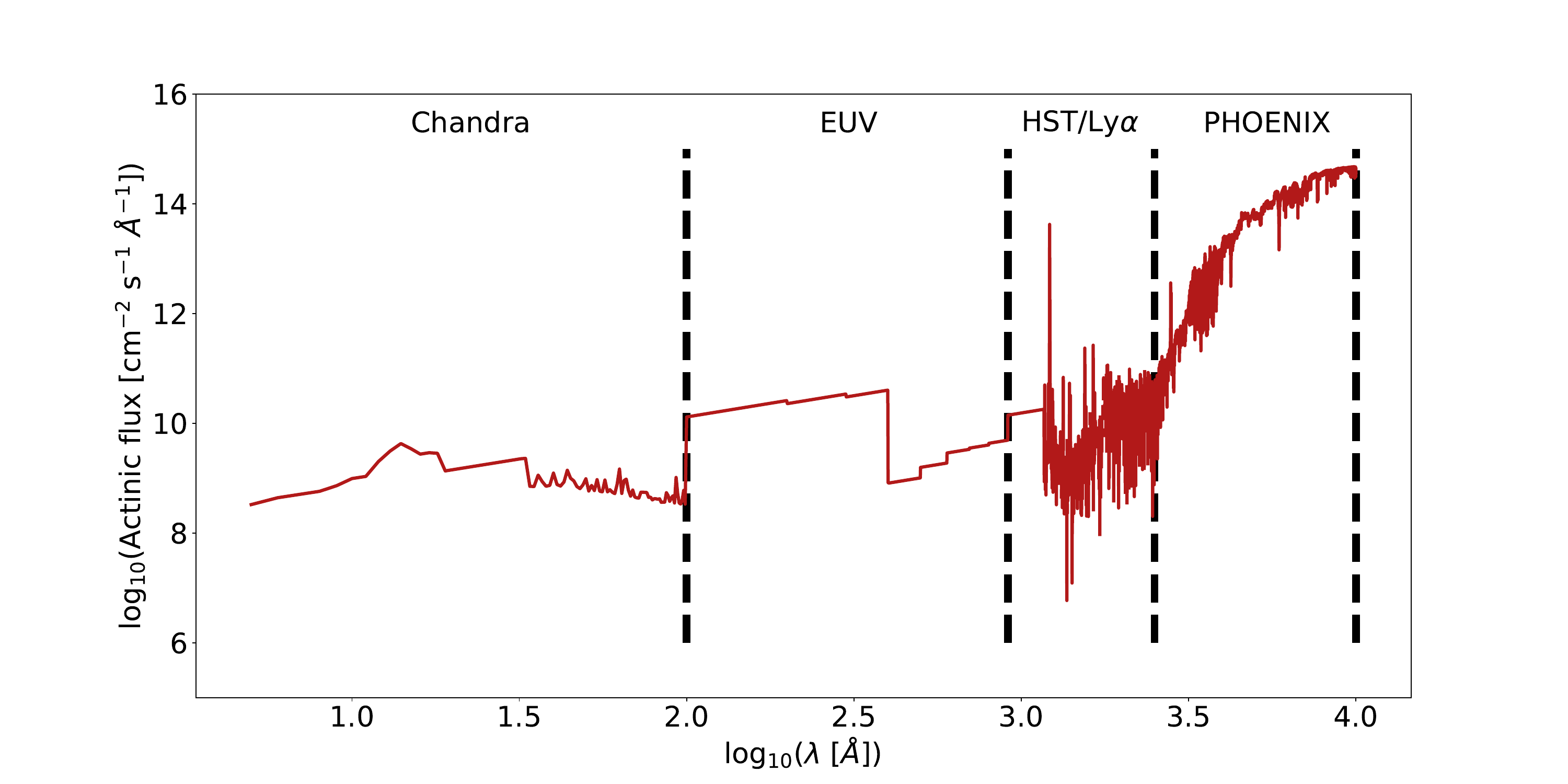}
        \caption{Input spectral energy distribution (SED) for host star expressed in actinic flux.}
        \label{fig:app_sed}
    \end{figure*}

   The input SED used for the photo-chemistry of ARGO is shown in Fig.~\ref{fig:app_sed}. The SED is obtained from the MUSCLES survey where the star GJ667C is chosen as the closest analogue to HATS-6. The SED is combined from different spectra achieved through observations and models, as indicated in the sections separated by the dashed lines (see e.g. \cite{france_muscles_2016} for details). The spectrum has been binned and adapted to the desired wavelength range 1-10,000 nm. 

\end{appendix}

\end{document}